\documentclass[10pt,twocolumn,letterpaper]{article}
\usepackage{authblk}
\usepackage[misc]{ifsym}
\usepackage[pagenumbers]{cvpr} 

\usepackage[accsupp]{axessibility}

\usepackage{graphicx}
\usepackage{booktabs}
\usepackage{enumitem}
\usepackage{arydshln} 

\usepackage{amsmath,amssymb,amsfonts}
\usepackage{algorithmic}
\usepackage{graphicx}
\usepackage{textcomp}
\usepackage{xcolor}
\usepackage{multirow}
\usepackage{mathrsfs}
\usepackage{booktabs}
\usepackage{color}
\usepackage{bm}
\usepackage[all, color]{xy}
\usepackage{pifont}
\usepackage{microtype}

\usepackage[pagebackref,breaklinks,colorlinks]{hyperref}

\usepackage[capitalize]{cleveref}
\crefname{section}{Sec.}{Secs.}
\Crefname{section}{Section}{Sections}
\Crefname{table}{Table}{Tables}
\crefname{table}{Tab.}{Tabs.}

\def\cvprPaperID{4925} 
\def\confName{CVPR}
\def\confYear{2023}

\makeatletter
\def\thanks#1{\protected@xdef\@thanks{\@thanks
        \protect\footnotetext{#1}}}
\makeatother

\begin{document}
\setlength{\textfloatsep}{5pt}  

\title{Bitstream-Corrupted JPEG Images are Restorable: Two-stage Compensation and Alignment Framework for Image Restoration\vspace{-1.5em}}


\author[1]{Wenyang Liu\thanks{$^\ast$Corresponding authors}}
\author[1$\ast$]{Yi Wang}
\author[1$\ast$]{Kim-Hui Yap}
\author[2]{Lap-Pui Chau\vspace{-0.8em}} 
\affil[1]{\textit {School of Electrical and Electronics Engineering, Nanyang Technological University, Singapore}}
\affil[2]{\textit {Dept. of Electronic and Information Engineering, The Hong Kong Polytechnic University, Hong Kong}}
\affil[ ]{\tt \small  {\{wenyang001, wang1241\}@e.ntu.edu.sg, ekhyap@ntu.edu.sg, lap-pui.chau@polyu.edu.hk}\vspace{-1.5em}}

\renewcommand\Authands{ and }

\maketitle




\begin{abstract}

In this paper, we study a real-world JPEG image restoration problem with bit errors on the encrypted bitstream. The bit errors bring unpredictable color casts and block shifts on decoded image contents, which cannot be resolved by existing image restoration methods mainly relying on pre-defined degradation models in the pixel domain. To address these challenges, we propose a robust JPEG decoder, followed by a two-stage compensation and alignment framework to restore bitstream-corrupted JPEG images. Specifically, the robust JPEG decoder adopts an error-resilient mechanism to decode the corrupted JPEG bitstream. The two-stage framework is composed of the self-compensation and alignment (SCA) stage and the guided-compensation and alignment (GCA) stage. The SCA adaptively performs block-wise image color compensation and alignment based on the estimated color and block offsets via image content similarity. The GCA leverages the extracted low-resolution thumbnail from the JPEG header to guide full-resolution pixel-wise image restoration in a coarse-to-fine manner. It is achieved by a coarse-guided pix2pix network and a refine-guided bi-directional Laplacian pyramid fusion network. We conduct experiments on three benchmarks with varying degrees of bit error rates. Experimental results and ablation studies demonstrate the superiority of our proposed method. The code will be released at \url{https://github.com/wenyang001/Two-ACIR}.

\end{abstract}

\vspace{-0.1in}
\section{Introduction}
\label{sec:intro}

\begin{figure}[t]
\centering
\includegraphics[width=3.2in]{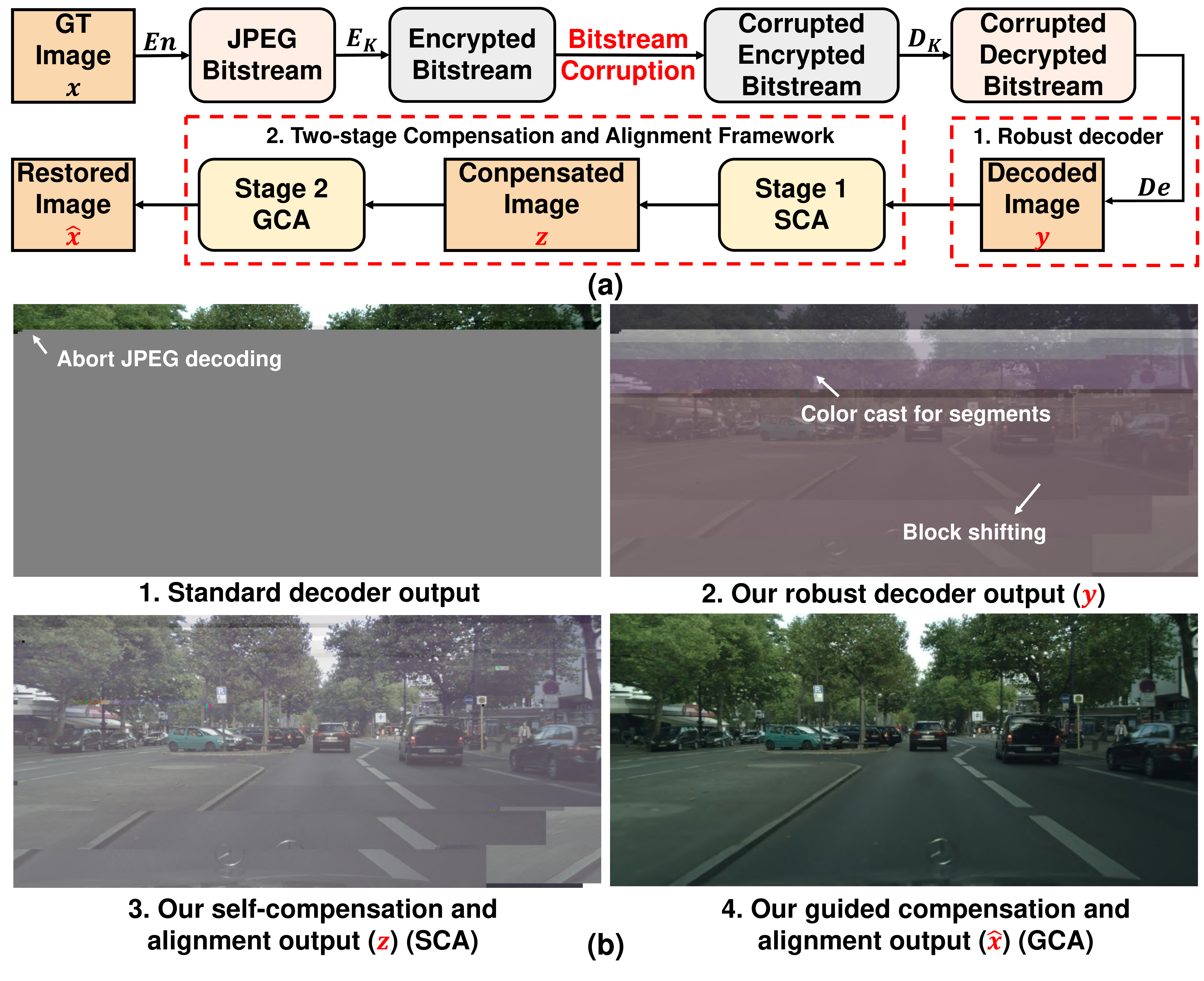}
\captionsetup{font=footnotesize}
\vspace{-0.15in}
\caption{(a) Our work considers a real-world JPEG image restoration problem with bit errors on the encrypted bitstream, where $En$/$De$ represent JPEG encoding/decoding and $E_K$/$D_K$ represent encryption/decryption employed in disks with the secret key $K$. We propose a robust JPEG decoder, followed by a two-stage compensation and alignment framework to address this problem. (b) Comparison of the standard decoder results with our robust decoder results and our proposed two-stage framework results. The proposed robust decoder can decode the corrupted JPEG bitstream and the proposed two-stage framework can ultimately restore high-quality images gradually from the decoded color-casted and misaligned images.}
\label{f:ss}
\end{figure}

Image restoration is a long-standing problem in computer vision that has been extensively studied. Given a degraded image, e.g., noisy, downscaled, hazing, or masked image, existing image restoration works in image deblur~\cite{pan2016blind, cho2021rethinking}, dehazing~\cite{qu2019enhanced}, inpainting~\cite{xie2019image, liu2018image}, superresolution (SR)~\cite{dong2015image, wang2020deep} are capable of restoring the high-quality counterpart, respectively. These methods are mainly based on pre-defined image degradation models in the pixel domain, but few attempts have been made in JPEG image restoration with the corrupted bitstream. The big challenge of bitstream-corrupted image restoration is the incurred JPEG decoding failures make the decoding process stop at the bit errors and the following bits cannot be decoded, as shown in Fig.~\ref{f:ss} (b1).

In the real world, bit errors occur naturally in JPEG bitstream stored in digital devices, and as memory cells wear out~\cite{tai2019s}, uncorrectable bit errors are exposed externally. NAND flash memory, as a type of non-volatile storage technology, is widely used in portable devices to store users' data. Due to technology trends, it exhibits progressively shorter lifetimes and increasingly relies on error correction codes (ECC) to ensure the data storage integrity~\cite{mielke2017reliability, pan2011exploiting}. It is well-known that~\cite{van2017nand, mielke2017reliability} raw bit error rate (RBER) of NAND flash memory grows rapidly as the program/erase cycle, temperature, and retention years increase. As a result, bit errors may exceed ECC's error correction capability and cause unrecoverable bit errors. In addition, if the storage device is severely damaged, or the ECC controller is not functioning correctly, standard data reading~\cite{van2017bit} may not be possible. Chip-off analysis~\cite{van2015mathematical} is often required to expose data in this case, but it may more likely result in unpredictable bit errors in the resolved data.

File carving~\cite{pal2009evolution} is an essential memory forensic technique that allows files to be recovered from unreliable NAND flash memory. While existing JPEG file carving methods~\cite{mohamad2011carving, garfinkel2007carving, uzun2015carving, uzun2019jpg} mainly focus on JPEG file carving in the absence of filesystem metadata, few consider the situation when the JPEG file itself is corrupted. Bit errors in the JPEG bitstream can severely deteriorate the decoded image quality by two kinds of error propagation~\cite{kuo2019long}. In addition, from Android 5.0, full-disk encryption (FDE)~\cite{khati2019full,khati2017full} is introduced to protect users' privacy. Once an Android device is encrypted, all user-created data will be automatically encrypted before committing it to disk and automatically decrypted before accessing it from disk. For encrypted files stored in an Android device, bit errors caused by the unreliable NAND flash memory are directly reflected on the encrypted data, making bit errors of the decrypted file become much more serious. This issue brings a significant challenge to existing works.

Recently, deep learning methods~\cite{zhang2019deep,wang2020deep, xie2019image, liu2018image} have shown great power in image restoration problems due to their powerful feature representation ability. However, existing image restoration methods may not be apt for the above-mentioned problem because of unpredictable color casts and block shifts of decoded image contents caused by bit errors. As Fig.~\ref{f:ss} (b1, b2) shows, decoders fail to generate visually consistent images that may not be directly used for the end-to-end training of existing image restoration methods.

Given the facts above, it is natural to raise a question: given a corrupted JPEG bitstream, is it possible to restore the image contents? With consideration of the FDE employed in smartphones for privacy, the damaged JPEG image $y$ in the pixel domain can be formulated as:

\vspace{-0.1in}
\begin{equation}
\label{e:Problem}
y = De(D_K(Bitflip(E_K(En(x))))
\end{equation}

\noindent where $x$ represents the initial JPEG image, $D_K$ and $E_K$ represent decryption and encryption of FDE, $De$ and $En$ represent JPEG decoding and encoding, $E_K(En(x))$ represents the corresponding encrypted JPEG bitstream by the secret key $K$, and $Bitflip$ represents random bit errors on the encrypted data. To simplify the problem, we assume the secret key is already known.

In this paper, we propose a robust JPEG decoder, followed by a two-stage compensation and alignment framework to restore bitstream-corrupted JPEG images. Specifically, the robust decoder adopts an error-resilient mechanism, which can decode the corrupted JPEG bitstream completely (see Fig.~\ref{f:ss} (b2)), compared to the aborting of JPEG decoding in the standard decoder (see Fig.~\ref{f:ss}(b1)). To further resolve the color cast and block shift problem in our decoded images, we propose a two-stage compensation and alignment framework, i.e., self-compensation and alignment (SCA) stage and guided-compensation and alignment (GCA) stage. In the first stage, SCA cast the problem as a segment detection problem and adaptively estimates suitable color and block offsets for each segment to perform block-wise image color compensation and alignment via image content similarity. In the second stage, GCA leverages the extracted low-resolution thumbnail (normally 160$\times$120~\cite{kee2010digital, parulski2018digital}) from the JPEG header to guide full-resolution pixel-wise image restoration. The GCA is achieved by coarse-to-fine neural networks, including a coarse-guided pix2pix network and a refine-guided bi-directional Laplacian pyramid fusion network. As Figs.~\ref{f:ss} (b3, b4) show, the proposed two-stage framework deals with the color cast and block shift problem and restores high-quality images ultimately. In summary, our contributions are as follows:

\begin{itemize}[itemsep=2pt,topsep=0pt,parsep=0pt]
\item To the best of our knowledge, this is the first work to restore the JPEG image with bit errors on the encrypted bitstream. Unlike existing works based on pre-defined degradation models in the pixel domain, the discussed problem in the bitstream domain causes unpredictable color casts and block shifts
on decoded images, which is challenging and of great practical value.

\item We propose a two-stage compensation and alignment scheme for this problem, where the SCA stage and GCA stage are proposed and combined into an end-to-end architecture. The SCA is based on image content similarity without training data, and the GCA employs the coarse-guided pix2pix network and the refine-guided bi-directional Laplacian pyramid fusion network to gradually restore full-resolution images.

\item Extensive experiments and ablation studies have been conducted to demonstrate the superiority of our proposed method. Even for 2k-resolution images, our proposed method can restore high-fidelity images with faithful details, achieving PSNR up to \textbf{38.92 dB} with \textbf{5.52 dB} significant improvement compared to the baseline EPDN method~\cite{qu2019enhanced}.

\end{itemize}

\begin{figure*}[htbp!]
\centering
\vspace{-0.2in}
\includegraphics[width=6.8in]{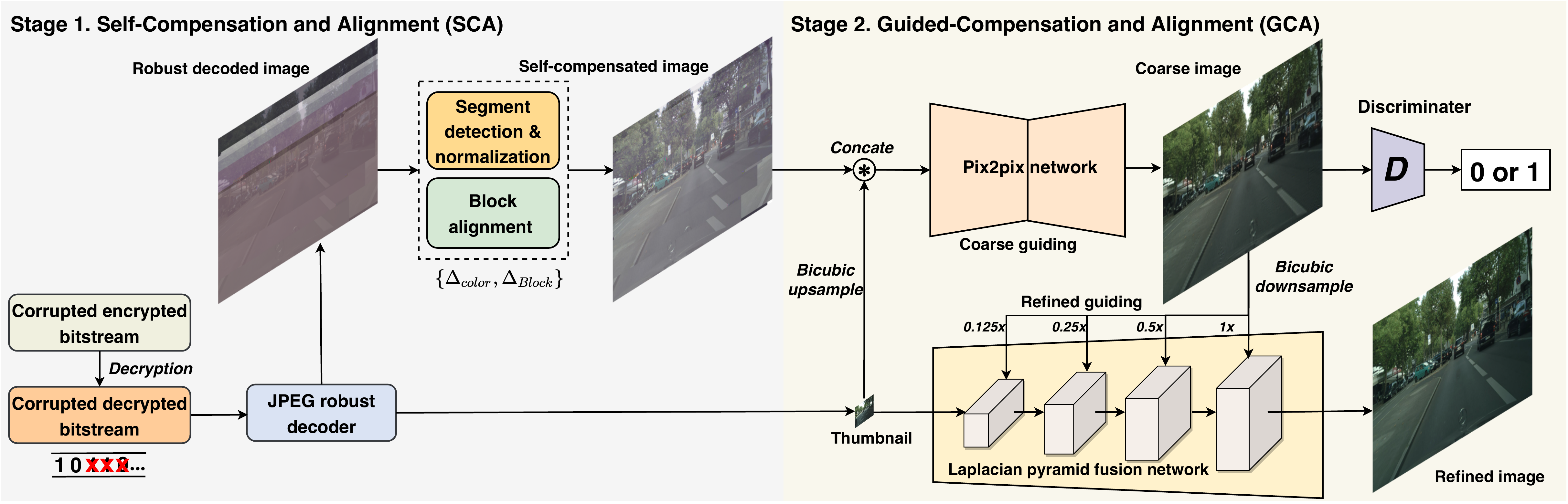}
\captionsetup{font=footnotesize}
\vspace{-0.1in}
\caption{Overall structure of our method with a robust decoder, followed by a two-stage alignment and compensation framework. The input is the JPEG corrupted encrypted bitstream. After the decryption, the JPEG corrupted decrypted bitstream is sent to the JPEG robust decoder to be fully decoded and to extract the thumbnail. In the first stage, SCA can adaptively perform block-wise image color compensation and alignment based on the estimated color and block offsets $\{\Delta_{Color}, \Delta_{Block}\}$. In the second stage, GCA leverages the extracted low-resolution thumbnail both in a coarse-guided pix2pix network and a refine-guided Laplacian pyramid fusion network to guide full-resolution pixel-wise image restoration in a coarse-to-fine manner.
} 
\label{f:network}
\vspace{-0.2in}
\end{figure*}

\vspace{-0.1in}
\section{Background and Related Work}
\vspace{-0.05in}
\textbf{JPEG structure.} In a standard JPEG encoding~\cite{marcellin2000overview}, an image is divided into multiple blocks of $8 \times 8$ pixels, where each block undergoes color space transform, discrete cosine transform (DCT), quantization, differential pulse code modulation coding (DPCM), run-length encoding (RLE), and Huffman coding. Since the employed Huffman coding is a variable-length coding method, bit errors in the encoded bitstream may lead to serious error propagations. A recent work~\cite{kuo2019long} points out two kinds of error propagations, i.e., bit error propagation and DC error propagation, which can severely deteriorate the decoded image quality. 

\begin{figure}[b]
\centering
\captionsetup{font=footnotesize}
\includegraphics[width=3.2in]{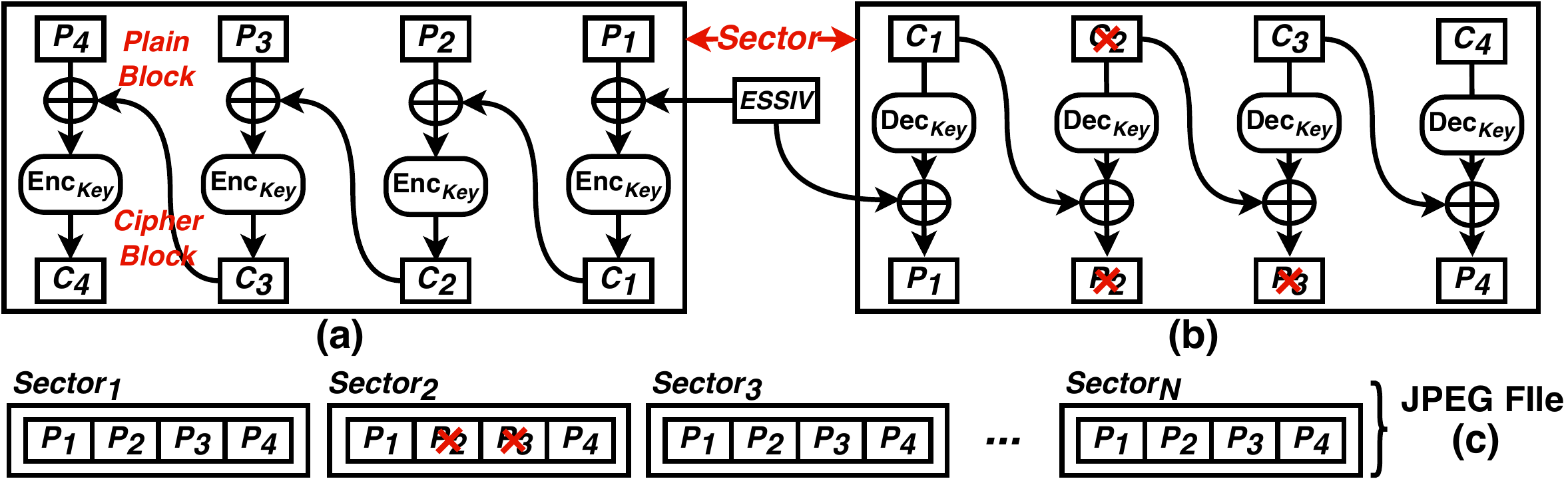}  
\vspace{-0.1in}
\caption{Encryption/decryption of a sector is sector-independent. (a) Encryption. (b) Decryption with bit errors in $c_2$. (c) The decrypted JPEG bitstream with the bit errors shown in (b) would not cause bit errors in other sectors.} 
\label{f:enc}
\end{figure}

\textbf{Full-disk encryption.} In storage devices, users' data are read and written in a sector-addressable device where each sector is typically 512 or 4,096 bytes. Full-disk encryption (FDE)~\cite{khati2017full} is a widely used cryptography method in smartphones that allows each sector of a disk volume to be encrypted independently to protect users' data privacy. There are several encryption modes in FDE, of which the cipher block chaining (CBC)~\cite{khati2017full} is the most widely used shown in Fig.~\ref{f:enc}. Taking the encryption for illustration, the plaintext of a sector is divided into blocks ($P_{i}$) of $n$ bits, typically 128, and each plain block $P_{i}$ is exclusive ORed (XORed) with the previous encrypted cipher block $C_{i-1}$, except for $P_{1}$ that is XORed with the encrypted salt-sector initialization vector (ESSIV) determined by the secret key and unique sector number. The corresponding output is encrypted by the Advanced Encryption Standard (AES) algorithm to obtain $C_{i}$. In this condition, a bit error occurring in a cipher block, e.g., $C_2$ of Sector 2, directly causes two adjacent plain blocks $P_2$, $P_3$ corrupted after decryption as Fig.~\ref{f:enc}(b) shows. Hence, bit errors appearing in encrypted data cause much more bit errors in decrypted data. Fortunately, both encryption and decryption are sector-independent, so bit errors in Sector 2 would not cause bit errors in other sectors as Fig.~\ref{f:enc}(c) shows.

\textbf{Image restoration.} Currently, a wider variety of image degradation models have been studied, e.g., image deblur~\cite{pan2016blind, cho2021rethinking}, image inpainting~\cite{xie2019image, liu2018image}, image superresolution~\cite{dong2015image, wang2020deep}, and image dehazing~~\cite{qu2019enhanced, li2018benchmarking}. Inspired by the huge success of Convolutional Neural Networks (CNNs) in image processing, CNN-based approaches have become mainstream and achieved different image-to-image translations. Dong \textit{et al.} proposed the first end-to-end CNN network called SRCNN to directly learn a non-linear mapping from low-resolution images to high-resolution images. After that, various network architectures have been proposed for super-resolution, e.g., deep residual learning~\cite{zhang2017beyond}, Laplacian pyramid learning~\cite{lai2018fast}. Apart from super-resolution, Cai \textit{et al.}~\cite{cai2016dehazenet} proposed an end-to-end CNN-based network for image dehazing called DehazeNat. Motivated by the success of cGAN~\cite{creswell2018generative}, Yanyun \textit{et al.}~\cite{qu2019enhanced} proposed a GAN-based pix2pixhd~\cite{isola2017image, wang2018high} model followed by a well-designed enhance blocks to learn a mapping from hazed images to dehazed images. These works mainly consider image restoration based on pre-defined image degradation models on in the pixel domain, but few consider JPEG bitstream corrupted image restoration.

\vspace{-0.1in}
\section{Our Method}
\vspace{-0.05in}
The overall structure of our model is shown in Fig.~\ref{f:network}, consisting of a robust JPEG decoder and a two-stage alignment and compensation framework, i.e., a self-compensation and alignment (SCA) stage and a guided-compensation and alignment (GCA) stage. Given a corrupted JPEG bitstream after decryption, it is first processed by the JPEG robust decoder to make the compressed image data fully decoded, and extract the thumbnail from the JPEG header. For the corrupted image, it is then sent to the SCA stage to adaptively perform block-wise image color compensation and alignment based on the estimated color and block offsets $\{\Delta_{Color}, \Delta_{Block}\}$. After that, GCA first leverages the bicubic upsampled thumbnail to coarsely guide the self-compensated image to achieve pixel-wise restoration through a pix2pix network, and then gradually fuses the low-resolution thumbnail with multi-scale bicubic downsampled images from the pix2pix network by the proposed Laplacian pyramid fusion network, restoring the final full-resolution refined image.

\vspace{-0.05in}
\subsection{Robust Decoder}
\begin{figure}[t]
\centering
\includegraphics[width=3.2in]{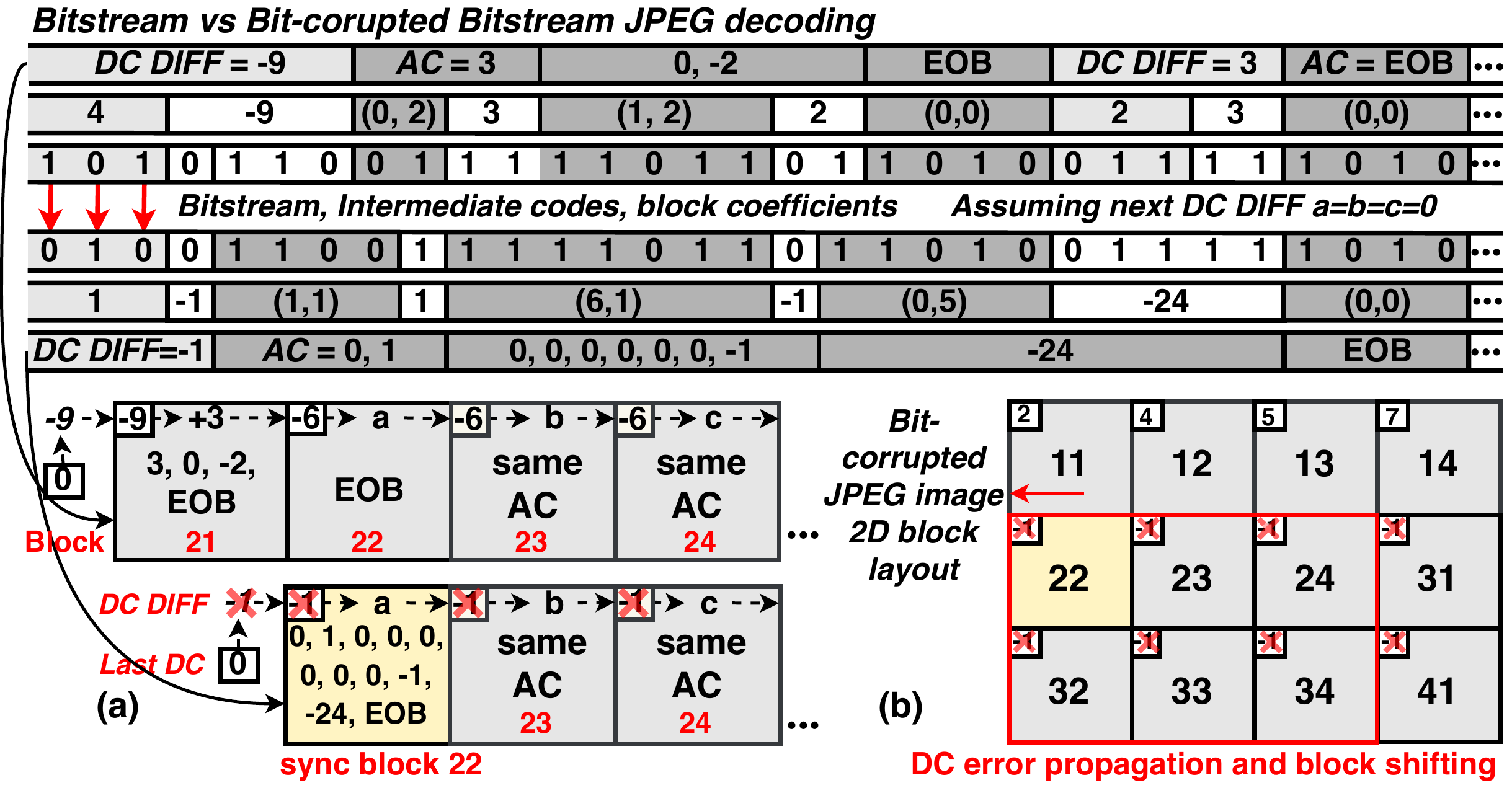}  
\vspace{-0.1in}
\captionsetup{font=footnotesize}
\caption{(a) Decoding comparison of the correct bitstream and a bit-corrupted bitstream whose first three bits are bit-flipped. The bit-corrupted bitstream achieves self-synchronization after block$_{22}$. (b) Two major problems are introduced after self-synchronization, i.e., DC error propagation and block shift.} 
\label{f:sych}
\vspace{-0.05in}
\end{figure}

\vspace{-0.05in}
\textbf{Self-synchronization.} Huffman coding is a variable-length encoding method that is widely used in JPEG to compress data. Recently, a published study~\cite{uzun2015carving} observed that the self-synchronization property of JPEG files seems to be possessed by JPEG files thanks to the large use of \textbf{\emph{EOB}}, a special codeword that is used to indicate the rest of decoded AC coefficients of a block are all zeros. Self-synchronization property means that bit errors in a bitstream can cause incorrect decoding at the beginning, but the decoder eventually can get the same decoding sequence as the original. As Fig.~\ref{f:sych}(a) shows, although there are three bits changed in the bitstream that is going to start block$_{21}$ decoding, the decoder still can re-synchronize at block$_{23}$. However, this property is not utilized by standard JPEG decoders. The core reason is that standard JPEG decoders lack error-resilient techniques. Once a decoding failure occurring in decoding, an exception is reported to abort the remaining blocks' decoding.

\begin{figure}[b]
\centering
\includegraphics[width=2.9in]{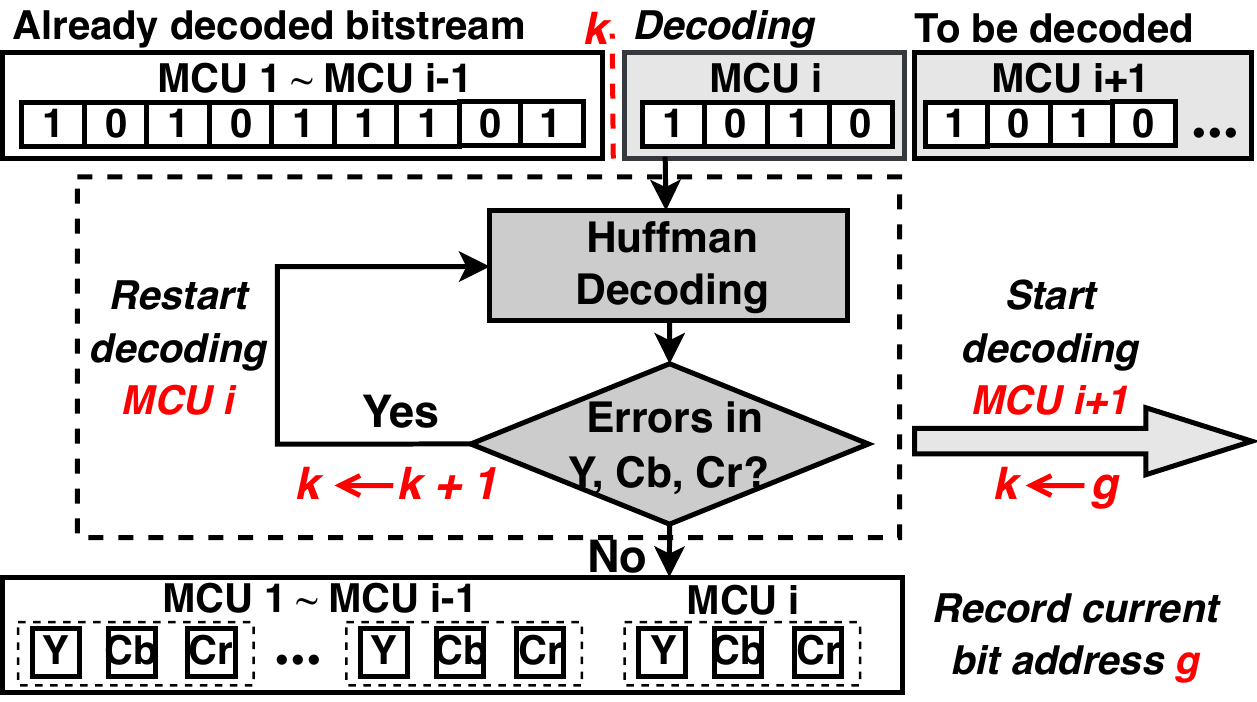}  
\vspace{-0.1in}
\captionsetup{font=footnotesize}
\caption{Error resilient mechanism in our robust JPEG decoder.}
\label{f:diag}
\end{figure}

\textbf{Robust JPEG decoding.} Here, we propose an error-resilient mechanism in our robust decoder shown in Fig.~\ref{f:diag}. There are two potential causes of a decoding failure that can be detected in our decoder when processing Huffman decoding in a Minimum Coded Unit (MCU) block. The first is when the decoder encounters an invalid codeword that cannot be found in the Huffman tables of the JPEG header. The second is when the number of decoded coefficients of an 8$\times$8 block is more than 64, called coefficients overflow. Once a decoding failure is detected, the error-resilient mechanism will discard a few bits to make the rest of the JPEG decoding can be continually processed, even though it may get some wrongly decoded blocks. As Fig.~\ref{f:diag} shows, assuming $\text{MCU}_1 \sim \text{MCU}_{i-1}$ are already correctly decoded and $\text{MCU}_i$ is being decoded, the proposed decoder start decoding at bit address $k$ to get $Y$, $Cb$, $Cr$ blocks of $\text{MCU}_i$, respectively. Once a decoding failure is detected in the whole $\text{MCU}_i$ decoding, our decoder will discard already decoded blocks of $\text{MCU}_i$ and restart decoding JPEG bitstream at bit address $k+1$ until $\text{MCU}$ can be fully decoded without any errors. After that, the decoder will save the decoded $\text{MCU}_i$, record the bit address $g$, and start the next $\text{MCU}_{i+1}$ decoding at bit address $g$. 

Although the proposed error-resilient mechanism can make JPEG files be decoded completely, the decoding results still have two problems as Fig.\ref{f:sych}(b) shows. The first problem is called DC error propagation. Since DPCM is employed to encode DC coefficients, the encoded DC value is the difference between the current block DC with the previous block DC. Therefore, although the remaining blocks' decoding is the same after block$_{22}$, the sync block$_{22}$ DC changing leads to the DC of these blocks being changed as Fig.\ref{f:sych}(b) shows. The second problem is called block shift. Since JPEG files construct 2D images by stacking blocks one by one from top to bottom, left to right, self-synchronization eats up the bitstream belonging to the block$_{21}$, and hence causes the remaining blocks to left shift one block.

\vspace{-0.05in}
\subsection{Self-Compensation and Alignment}
\vspace{-0.05in}
\textbf{Segment detection and normalization.} In JPEG images, the DC coefficient of a block represents the average intensity of the 8$\times$8 pixels. A small DC coefficient variation $\Delta_{DC}$ can shift all 64 pixels of a block by $L \cdot \Delta_{DC}$, where $L$ is a constant value. The high value of the DC variation undoubtedly causes the pixels to exceed the specified pixel range [0, 255]. Observing that the DC error propagation results in the same pixel shift for the following consecutively and correctly decoded blocks (namely, blocks segment) before the next self-synchronization, we intuitively cast this problem as an image segment detection problem. Blocks inside a segment share the same DC shift and hence have smooth image contents after decoding, and blocks between two consecutive segments have a big difference in image contents after decoding. Taking advantage of this feature, we propose a segment detection method based on the 2D image content similarity. A segment point is deemed to be detected when image contents have an abrupt change. Each segmented point is determined by a horizontal and a vertical coordinate as $s = (h, v)$ in a 2D image, where $h, v \bmod 8 = 0$. We first detect $h$ and then detect $v$ by fixing $h$.

\begin{figure}[t]
\centering
\includegraphics[width=3.22in]{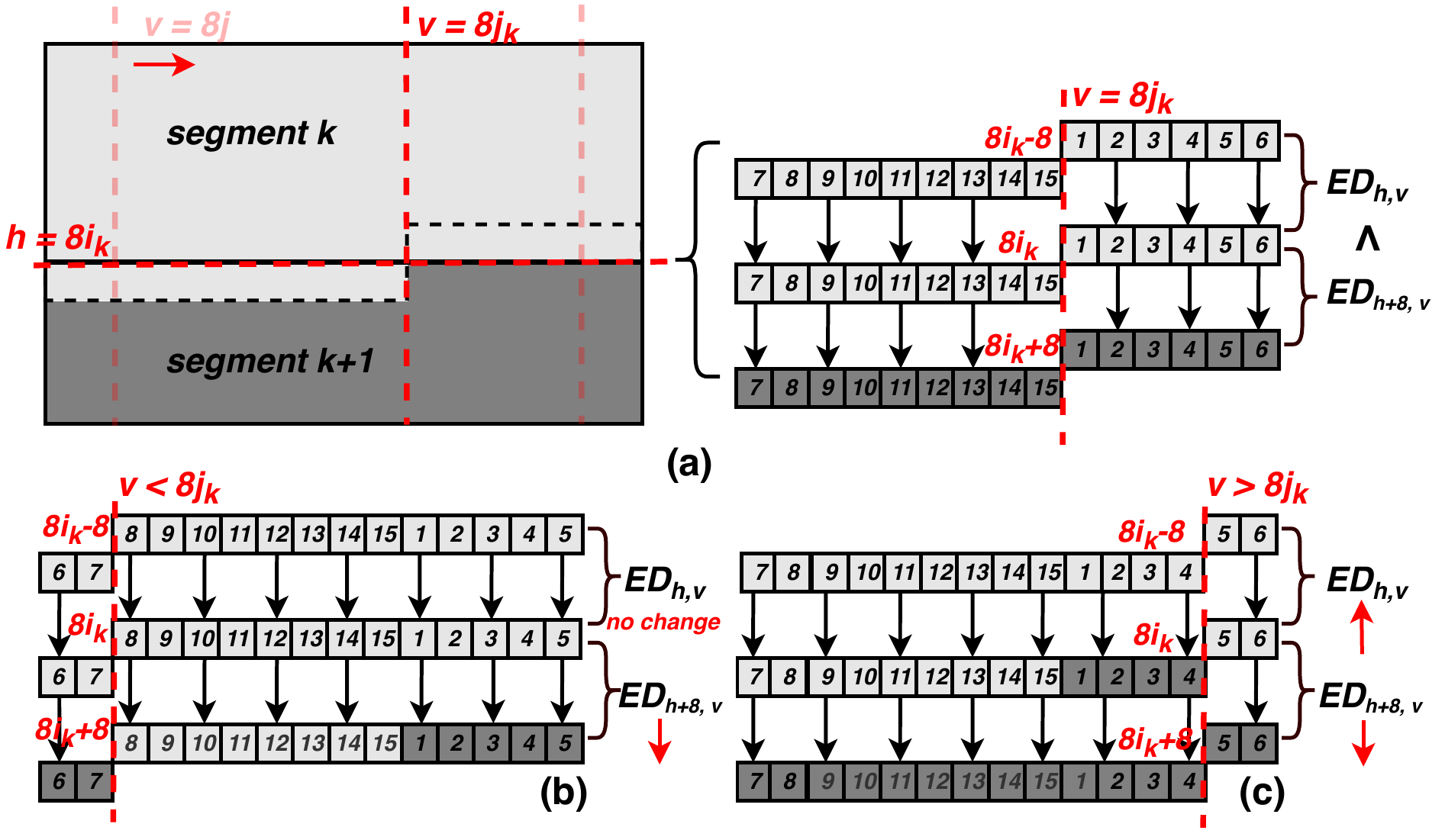}  
\vspace{-0.12in}
\captionsetup{font=footnotesize}
\caption{Vertical coordinate detection of the segmented point by calculating the coherence of ED (CED). The vertical coordinate $v$ is determined when CED reaches the maximum at $8j_k$.} 
\label{f:CED}
\vspace{-0.07in}
\end{figure}


For the horizontal coordinate $h$, it can be easily determined by checking the pixel similarity across the horizontal boundary of consecutive row blocks. Sum of Differences (SoD) and Euclidean Distance (ED) are two commonly used similarity metrics~\cite{tang2016recovery}. Lower SoD/ED values mean higher pixel similarity. Here, we select ED as a similarity metric to measure the pixel similarity across the horizontal boundary of consecutive row blocks, expressed as:

\begin{footnotesize}
\begin{equation}\label{e:ED}
ED_h=\frac{1}{W}(\sum_{i=1}^{W} \left(p_{h+1, i}-p_{h, i}\right)^{2})^{1/2},
\end{equation}
\end{footnotesize}


\noindent where $ED_h$ represents the pixel similarity between row $h$ and row $h-1$, $p_{h, i}$ is the corresponding RGB values of the pixels of the image at the position $(h, i)$, and $W$ is the width of the image. Once we calculate all EDs for each $h$, those having big ED are regarded as the horizontal coordinates of the candidate segmented points.






After the horizontal coordinate $h=8i_k$ of a segmented point is detected, the corresponding vertical coordinate $v=8j_k$ is required to be detected as Fig.~\ref{f:CED}(a) shows. A naive idea is to check the pixel similarity across the vertical boundary of consecutive blocks at $h=8i_k$. However, it is easier to be misled since it only considers 8 pixels difference across the vertical boundary. To address this problem, we propose a new vertical detection method based on the coherence of ED (CED)~\cite{tang2016recovery}. CED is defined as the difference between adjacent EDs of row pixel blocks, expressed as:


\begin{footnotesize}
\begin{gather}
\label{e:ED}
CED_{h, v} = |ED_{h+8, v} - ED_{h, v}|, \\
ED_{h, v} = \frac{1}{W} \left(\sum_{i=v}^{W} (p_{h+1, i}-p_{h, i})^{2} + \sum_{i=1}^{v} (p_{h+9, i}-p_{h+8, i})^2\right)^{1/2},
\end{gather}
\end{footnotesize}

\noindent where $CED_{h, v}$ represents the pixel similarity at the given segmented point $(h,v)$ and unlike $ED_{h}$ calculation, $ED_{h, v}$ is calculated by two parts addition according to the vertical coordinate $v$. Assuming two segments are divided by the segmented point $s_k = (8i_k, 8j_k)$ in Fig.~\ref{f:CED}(a) where the horizontal coordinate $h$ is already determined, to detect the vertical coordinate $v$, we calculate CEDs for all points $(h=8i_k, v=8j)$ for $j = 0,1,2...$. Since lower ED values mean higher pixel similarity, when $v < 8j_k$ shown in Fig.~\ref{f:CED}(b), compared to $v = 8j_k$, the $ED_{h+8, v}$ decreases because adjacent row blocks are more similar and the $ED_{h, v}$ is almost no changed because adjacent row blocks still belong to the same segment, resulting in the overall CED decreasing. And when $v > 8j_k$ shown in Fig.~\ref{f:CED}(c), the overall CED decreases due to the same reason. It can be seen that the vertical coordinate of the segmented point is determined when CED reaches the maximum at $v = 8j_k$.


After all segment points are detected, pixel normalization is performed on each segment. At each color channel, it first centers all pixels by subtracting the mean pixel value of the segment to compensate for the abnormal DC shift of each segment. To do so, most pixels are centered at zero with only a few isolated pixel values of sync blocks that are extremely high or low due to self-synchronization. These isolated pixel values stretch the real pixel range and are required to eliminate. We then use the \textbf{\emph{Clip}} function to ensure the overall pixel values in [-150, 150]. Finally, the min-max normalization is applied to the whole image to re-scale the pixel range back to [0, 255] for image display.


\textbf{Block alignment.} To ease the block shift problem caused by self-synchronization, we add an additional block alignment processing. Given a misaligned image, the alignment processing is to determine how many blocks each row should shift to align with the upper row. For each row $h$, each time the row shifts a block left or right, the corresponding $ED_h$ is calculated. The number of blocks required to be shifted for a row is determined when the $ED_h$ reaches the minimum. This alignment operation will continue until the last row of the image is aligned with the upper row.


\vspace{-0.05in}
\subsection{Guided-Compensation and Alignment}
\vspace{-0.05in}



\textbf{Thumbnail integration.} For most JPEG photos created by phones or digital cameras, the thumbnail is auto-created and embedded into the JPEG header of APP0 marker segment~\cite{wallace1991jpeg, kee2010digital}, which are stored separately versus the actual compressed data. Since the thumbnail is very small (typically 160$\times$120~\cite{kee2010digital, parulski2018digital})) compared to the actual compressed image data, it is more likely to escape random bit errors. Therefore, we introduce the thumbnail as the network additional input to guide the image reconstruction. The introduced thumbnail is first bicubic upsampled and then concatenated with the self-compensated image as the inputs of pix2pix network~\cite{qu2019enhanced}. Although the blurred thumbnail lacks details, it can bring the required color and block alignment information.

\textbf{Pix2pix network.} The pix2pix network is based on the previous works~\cite{isola2017image, qu2019enhanced}, consisting of multi-resolution generators and multi-resolution discriminators. 
In our paper, the pix2pix2 network concatenates the thumbnails and self-compensated images from the SCA to coarsely guide the resulting images with more consistent color and aligned textures. The details of the network can be seen in the Supplementary Materials. 

\begin{figure}[t]
\centering
\includegraphics[width=3.1in]{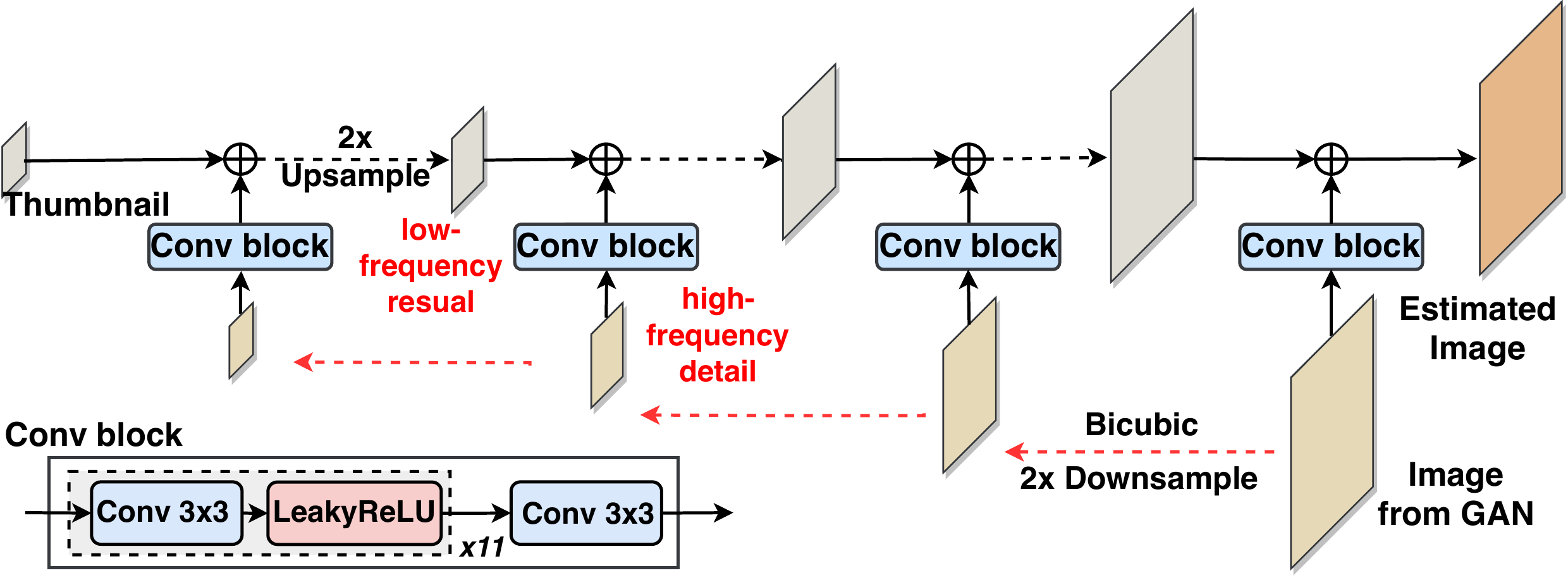}  
\vspace{-0.1in}
\captionsetup{font=footnotesize}
\caption{The structure of bi-directional Laplacian fusion network. Black dotted arrows indicate de-convolutional layers for upsampling and red dotted arrows indicate downsampling by bicubic interpolation. The convolution block is composed of 11 convolutional layers of a 3$\times$3 convolution and a Leaky Relu activation, plus one more 3$\times$3 convolutional layer. } 
\vspace{-0.05in}
\label{f:comp}
\end{figure}

\textbf{Laplacian pyramid fusion network.} To refine the coarse image from the pix2pix network, a pooling-enhanced module is used in EPDN~\cite{qu2019enhanced} to integrate the pix2pix network input, i.e., the thumbnail and the self-compensated image. However, observing that the thumbnail is blurred and the self-compensated image is not fully aligned, neither of them is a good choice for the pooling-enhanced module as they can make the network focus unrelated features for fusion, e.g., the upsampled thumbnail makes the network results blur. Conversely, instead of refining the coarse image, the proposed bi-directional Laplacian pyramid fusion network aims to refine the thumbnail gradually under the different scales of the coarse image guidance, as Fig.~\ref{f:comp}(b) shows. Unlike the Laplacian pyramid structure from existing works~\cite{lai2017deep, lai2018fast}, our structure is bi-directional and relies on both upsampling and downsampling processes. At first, from right to left, the coarse image is bicubic downsampled to generate images in different scales, which are then convolved with a convolution block to generate pyramid high-frequency features (gradually finer details). From left to right, the thumbnail as low-frequency residual is gradually element-wise added pyramid high-frequency details to get the final output. At each step, the element-wise added feature is 2x upsampled by a trainable de-convolution layer.






\begin{figure*}[htbp!]
\centering
\includegraphics[width=6.8in]{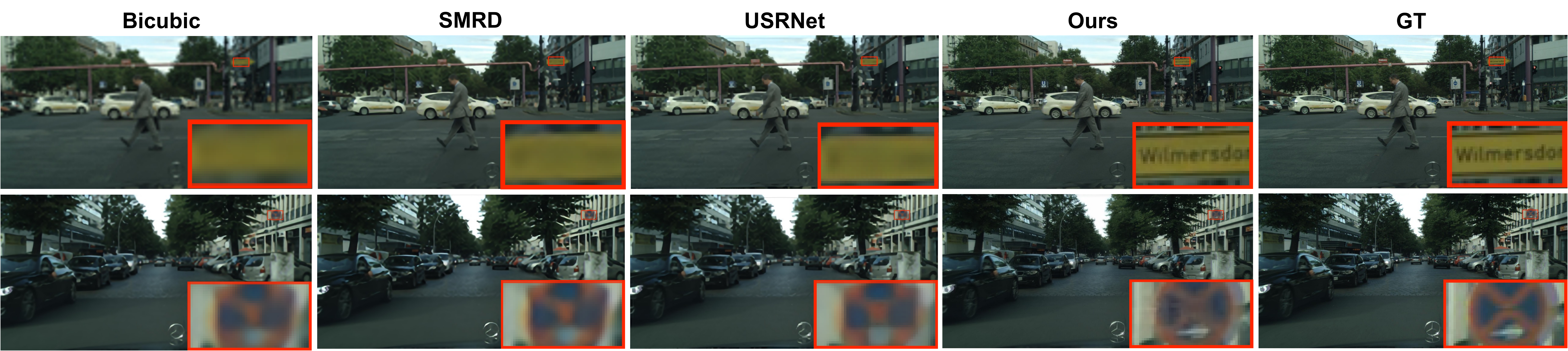}  
\captionsetup{font=footnotesize}
\vspace{-0.1in}
\caption{Visual comparison with bicubic and SOTA super-resolution methods for low-resolution thumbnails at scale factor 6.} 
\vspace{-0.25in}
\label{f:SR}
\end{figure*}

\textbf{Loss Function.} In this paper, we follow the same loss functions from EPDN~\cite{qu2019enhanced}, including the adversarial loss $L_A$, the feature matching loss $L_{FM}$, the perceptual loss $L_{VGG}$. Considering the employed Laplacian structure, we adopt a robust Charbonnier loss function~\cite{lai2017deep} $L_C$ to replace the $\ell_2$ fidelity loss, i.e., 

\begin{footnotesize}
\begin{equation}
      L_{C} = \frac{1}{L}\sum_{i=1}^{L} \left((\hat{X}_{i} - X_i)^2 + \epsilon^2) \right) ^ {1/2}, \\
\end{equation}
\end{footnotesize}

\noindent where $\hat{X_i}$ and $X_i$ denote the $i$-layer of the pyramid output and the ground truth, $L$ is the number of levels of the pyramid outputs, and $\epsilon$ is empirically set to $1\text{e-}3$. To make the network focus on the block alignment information learning, we add an additional edge loss $L_E$~\cite{lu2021single}, expressed as $L_{E} = \Vert S(\hat{X}) - S(X) \Vert_2$, where $S(\hat{X})$ and $S(X)$ denote the Sobel Operator on the estimated network output and the ground truth. The overall loss function is:

\begin{footnotesize}
\begin{equation}
     L = L_{A} + \lambda_1 L_{FM} + \lambda_1 L_{VGG} + \lambda_2 L_{E} + \lambda_3 L_{C}, 
\end{equation}
\end{footnotesize}

\noindent where $\lambda_1, \lambda_2, \lambda_3$ are user defined hyper-parameters. We follow the same alternative iteration training scheme~\cite{qu2019enhanced}, where the GAN module (generator and discriminator) is first optimized by the $L_A$, $L_{FM}$ and $L_{E}$, and the Laplacian pyramid fusion network and the generator is optimized by $L_{VGG}$, $L_{E}$, and $L_C$ in one training step. The generator's weights are updated twice during each training step. More details of the loss function can be seen in the Supplementary Materials.

\vspace{-0.1in}
\section{Experiment}
\vspace{-0.05in}
\subsection{Implementation Settings}
\vspace{-0.05in}

\textbf{Datasets.} We conduct extensive comparisons and ablation studies on AFHQ~\cite{choi2020stargan}, CelebA-HQ~\cite{liu2015faceattributes} and Cityscapes~\cite{cordts2016cityscapes}, corresponding to the image resolution 512$\times$512, 1024$\times$1024 and 2048$\times$1024, respectively. JPEG files from \cite{choi2020stargan,liu2015faceattributes,cordts2016cityscapes} are first encrypted by the full-disk encryption (FDE)~\cite{khati2019full,khati2017full} method, followed a bit error rates (BER) setting of $10^{-5}$ in~\cite{kuo2019long, van2017nand} on the encrypted bitstream, and then decrypted by the same FDE method to construct the datasets. As for the thumbnail setting, we assume that the thumbnail comes from the bicubic downsampling, and its maximum side of the thumbnail is fixed at 160~\cite{marcellin2000overview}.



\textbf{Training details.} We adopt Adam optimizer with a batch size of 4, a learning rate of 0.0002, and the exponential decay rates of $(\beta_1, \beta_2)=(0.6, 0.999)$ after epoch 100. The total epochs are set to 200. The corresponding hyper-parameters of the loss function are set as $(\lambda_1, \lambda_2, \lambda_3) = (10, 15, 150)$. We implement our model with the PyTorch on an NVIDIA GPU GeForce RTX 3090. 

\textbf{Evaluation Metrics.} The peak signal-to-noise ratio (PSNR) and structural similarity index (SSIM) in the $Y$ channel are used to measure the quality of restored images.

\vspace{-0.05in}
\subsection{Experimental Results} 
\vspace{-0.05in}
\textbf{Quantitative results.} We conduct experiments on the Cityscapes~\cite{cordts2016cityscapes} dataset. To evaluate our proposed method in handling different resolutions of JPEG image restoration, we use bicubic downsampling methods to get additional Cityscapes datasets with 1024$\times$512 and 512$\times$256 resolution. Tab.~\ref{t:Results_City} shows quantitative results with different methods. Note that since the standard decoder aborts JPEG decoding for the corrupted bitstream, we only show the PSNR and SSIM of our robust decoder here. It can be seen that our SCA can improve PSNR and SSIM by around 2 dB and 0.1. Based on the SCA's results, we compare our GCA network with a baseline image-to-image translation network EPDN~\cite{qu2019enhanced} which adopts the same pix2pix network. It shows our GCA outperforms EPDN by a significant margin, e.g., gaining PSNR of \textbf{8.54 dB} improvement in 512$\times$256 image reconstruction. The superior results of Cityscapes in different resolutions demonstrate the superiority of our proposed two-stage SCA and GCA method.

\begin{table}[t]
\centering
\captionsetup{font=footnotesize}
\caption{Quantitative comparison of different methods on different scales of Cityscapes~\cite{cordts2016cityscapes} datasets in terms of PSNR/SSIM.} 
\vspace{-0.1in}
\label{t:Results_City}
\footnotesize
\resizebox{\linewidth}{!}{
\begin{tabular}{lccc} \hline 
\multirow{2}{*}{Method}  & \multicolumn{3}{c}{Cityscapes~\cite{cordts2016cityscapes}} \\ \cline{2-4}
             & 512$\times$256 &  \ 1024$\times$512 &    2048$\times$1024 \\  
\hline 
 Robust decoder (Ours) & 12.41/0.58  & 12.38/0.64 & 12.37/0.70 \\   \hdashline
 SCA (Ours) &  14.34/0.72  &  14.32/0.75  &  14.31/0.78  \\ 
 SCA + EPDN &  33.51/0.94 &  33.25/0.94 & 33.40/0.95 \\  
 SCA + GCA (Ours) & \textbf{42.05/0.98}  & \textbf{40.09/0.97} &  \textbf{38.92/0.97}
 \\  
\hline          
\end{tabular}
}
\vspace{-0.15in}
\end{table}

\begin{figure}[t]
\centering
\includegraphics[width=3.2in]{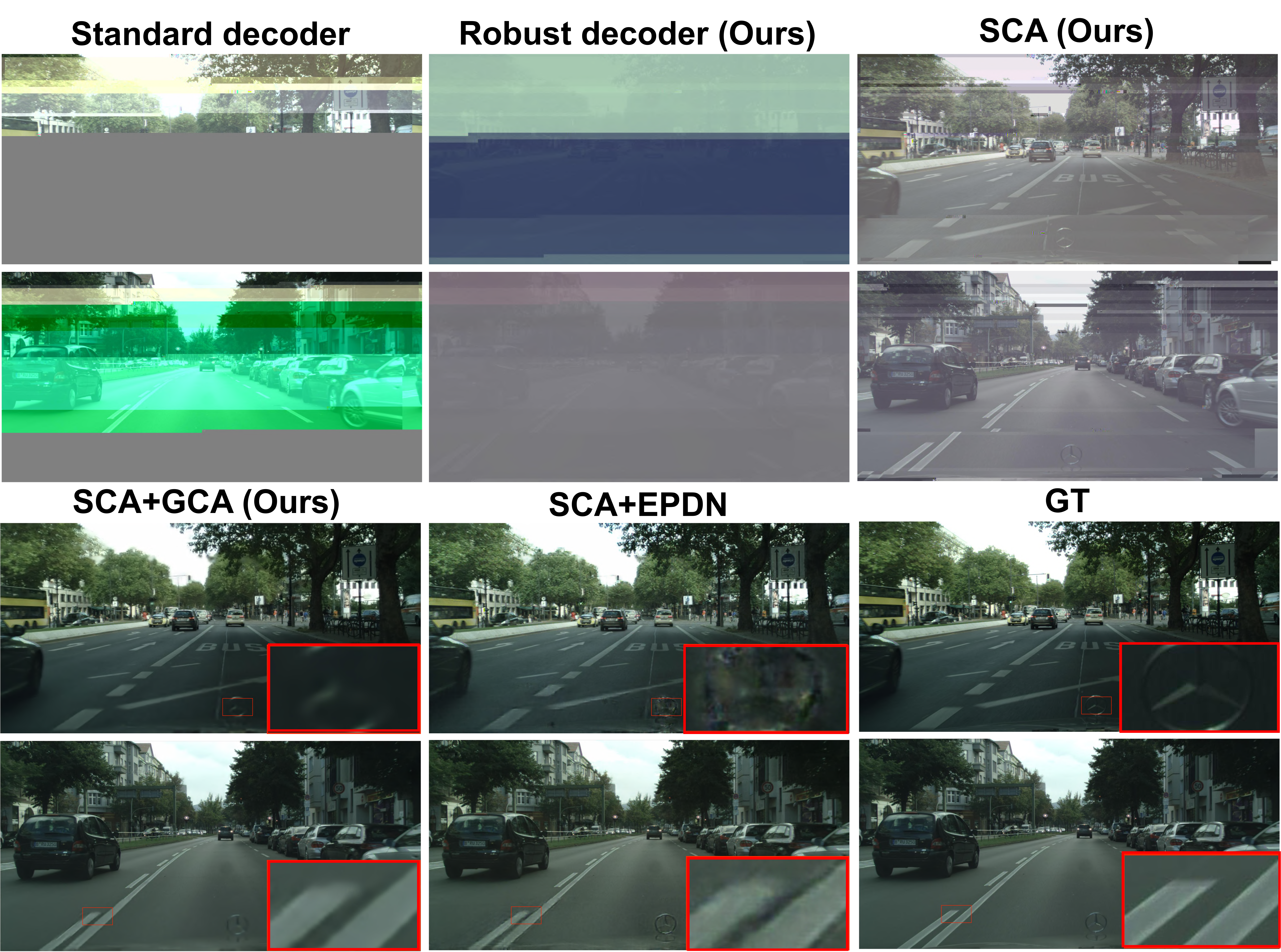}  
\vspace{-0.05in}
\captionsetup{font=footnotesize}
\caption{Visual comparison of our method with the standard decoder and the existing EPDN~\cite{qu2019enhanced} method. } 
\label{f:stage}
\end{figure}


\textbf{Qualitative results.} We qualitatively compare our method with different methods in Fig.~\ref{f:stage}. The standard decoder fails to decode the corrupted JPEG bitstream. The proposed robust decoder uses an error-resilient mechanism that makes the corrupted JPEG bitstream fully decoded but causes serious color casts and block shifts. Our proposed SCA adaptively compensates for the color and block offsets and hence delivers better visual results, demonstrating its effectiveness. Compared with EPDN refining the pix2pix network output, the proposed GCA refining the thumbnail gradually benefits from the coarse guiding and refined guiding, and hence delivers much-aligned and better visual image contents. We provide more visual results in the Supplementary Materials.

\begin{figure*}[htbp!]
\centering
\includegraphics[width=6.8in]{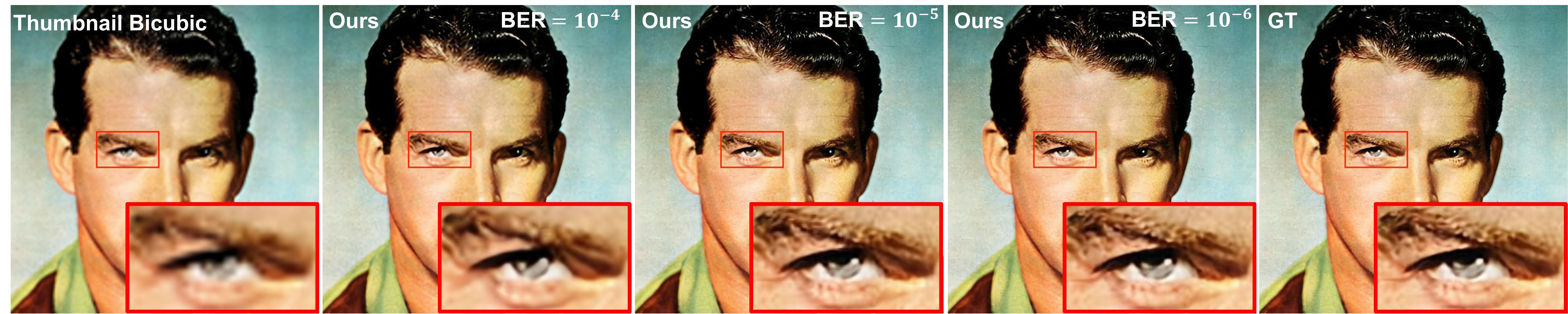}  
\captionsetup{font=footnotesize}
\vspace{-0.05in}
\caption{Visual comparison of our GCA results on varying BERs. The bicubic upsampled thumbnail and the ground truth are given for comparison.} 
\vspace{-0.25in}
\label{f:BER_detail}
\end{figure*}

\vspace{-0.05in}
\subsection{Ablation Study}
\vspace{-0.05in}
\textbf{Impact of proposed components.} We ablate several proposed components in the Cityscape dataset in Tab.~\ref{t:diff} as followings: 1) \textbf{w/o SCA}: remove the whole SCA (i.e., Segment detection \& normalization, block alignment). 2) \textbf{w/o Block alignment in SCA.} 3) \textbf{w/o Laplacian fusion in GCA}: remove the Laplacian pyramid fusion network in the GCA and use the EPDN's pooling-enhanced module instead. 4) \textbf{w/o Edge \& Charbonnier loss}: use the original $\ell_2$ fidelity instead of loss $L_C$ and $L_E$. For each setting, we will re-train our model. These ablation studies demonstrate that the proposed components are effective for the bitstream-corrupted JPEG image restoration, especially the proposed SCA and Laplacian fusion playing the most important role in the final restored image's quality.


\begin{table}[t]
\centering
\captionsetup{font=footnotesize}
\caption{Ablation Study of the impact of our proposed components in SCA and GCA stages.} 
\label{t:diff}
\footnotesize
\vspace{-0.1in}
\begin{tabular}{lccc}
\hline \multirow{2}{*}{Method} & \multicolumn{2}{c}{Cityscapes~\cite{cordts2016cityscapes}} \\  \cline{2-3}
          & 512$\times$256 & 1024$\times$512 \\
\hline 
w/o SCA  &  33.88/0.92 & 33.52/0.91 \\
w/o Block alignment in SCA  & 36.50/0.95     & 35.01/0.92  \\  \hdashline
w/o Laplacian fusion in GCA &  36.65/0.95 & 36.16/0.93   \\ 
w/o Edge \& Charbonnier loss &  41.27/0.98  & 39.53/0.97 \\   \hdashline   
Ours  &   \textbf{42.05/0.98} & \textbf{40.09/0.97}   \\ \hline    
\end{tabular}
\end{table}


\textbf{Impact of thumbnail-guided image restoration.} 
Current image restoration methods cannot resolve bitstream-corrupted JPEG files. Given recovered low-resolution thumbnails, we compare our thumbnail-guided method with state-of-the-art (SOTA) super-resolution (SR) methods in bicubic degradation on the Cityscape dataset. Because SR methods only support integer scale factors, we keep the low-resolution thumbnails at 170$\times$85. Results and qualitative comparisons of $\times 6$ scale are shown in Tab.~\ref{t:DeResults_S} and Fig.~\ref{f:SR}, respectively. Our method outperforms USRNet~\cite{zhang2020deep} and SRMD~\cite{zhang2018learning} in PSNR and SSIM, especially in the scale factor x6 with around PSNR/SSIM of \textbf{8 dB/0.1} improvement.

\begin{figure}[t]
\centering
\includegraphics[width=3.2in]{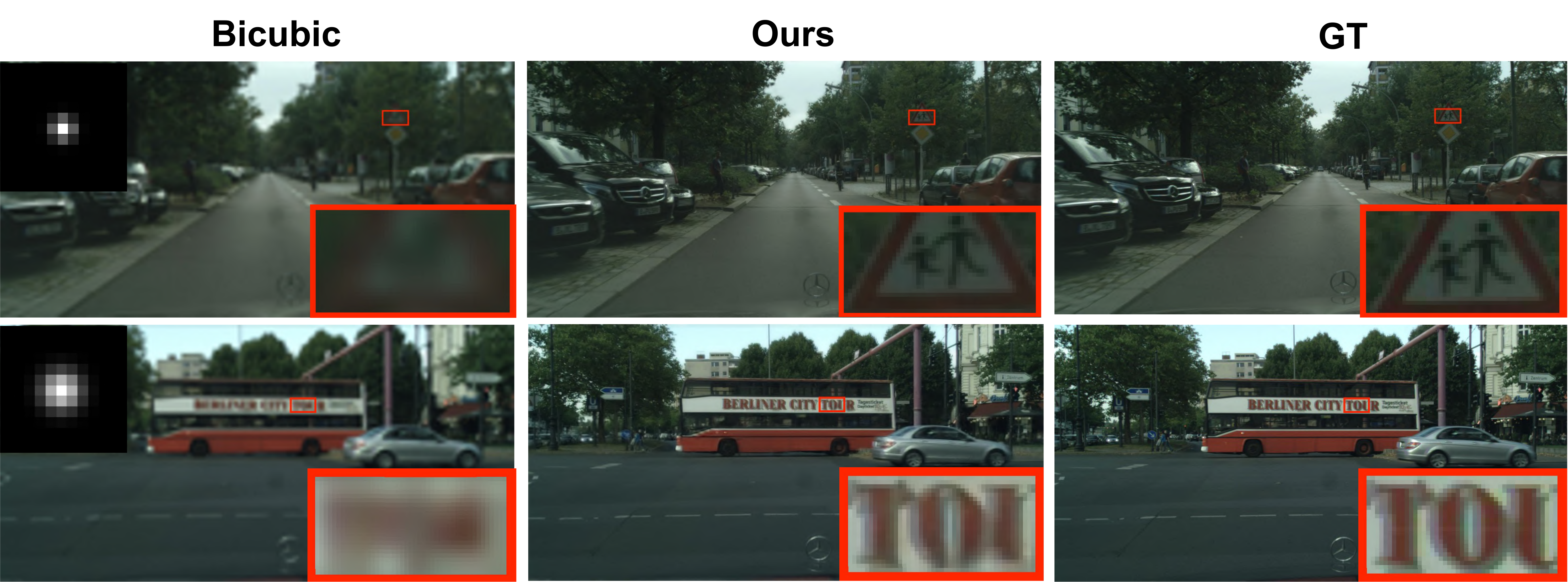}  
\vspace{-0.05in}
\captionsetup{font=footnotesize}
\caption{Visual comparison on unseen degraded thumbnails by isotropic Gaussian kernels with kernel widths ~\cite{zhang2020deep,zhang2018learning} of 0.7 (top) and 1.2 (bottom).}
\label{f:BER_b}
\end{figure}

\begin{table}[t]
\centering
\captionsetup{font=footnotesize}
\vspace{-0.1in}
\caption{Quantitative comparison in PSNR/SSIM of bicubic, SOTA SR, and our method at different scale factors on Cityscapes~\cite{cordts2016cityscapes} datasets}
\vspace{-0.1in}
\label{t:DeResults_S}
\footnotesize
\begin{tabular}{lcc}
\hline \multirow{2}{*}{Method}   & \multicolumn{2}{c}{Scale factors} \\   \cline{2-3}
       & $\times$3   &  $\times$6  \\
\hline 
 Bicubic  &  30.94/0.88 & 29.79/0.82 \\
 SRMD~\cite{zhang2018learning}         & 33.37/0.92 & 31.66/0.87   \\
 USRNet~\cite{zhang2020deep}  &     32.78/0.92  & 31.18/0.86      \\ 
 Ours   &   \textbf{42.05/0.98}  & \textbf{40.09/0.97}       \\  \hline
\end{tabular}
\end{table}


\vspace{-0.05in}
\subsection{Generalization Capability}
\vspace{-0.05in}
\textbf{Varying degradation of thumbnails.} We test the generalization of our method in handling different unseen degraded thumbnails, including a representative degradation in superresolution: isotropic Gaussian kernels with widths~\cite{zhang2020deep,zhang2018learning} of 0.7 and 1.2. The training of our GCA is under the bicubic degraded thumbnails, and the testing is under unseen degraded thumbnails. Qualitative comparisons are shown in Fig.~\ref{f:BER_b}. The results show that our method still can deliver impressive results for these unseen  thumbnails, which demonstrates its superiority in generalizability.



\begin{figure}[b]
\centering
\includegraphics[width=3.2in]{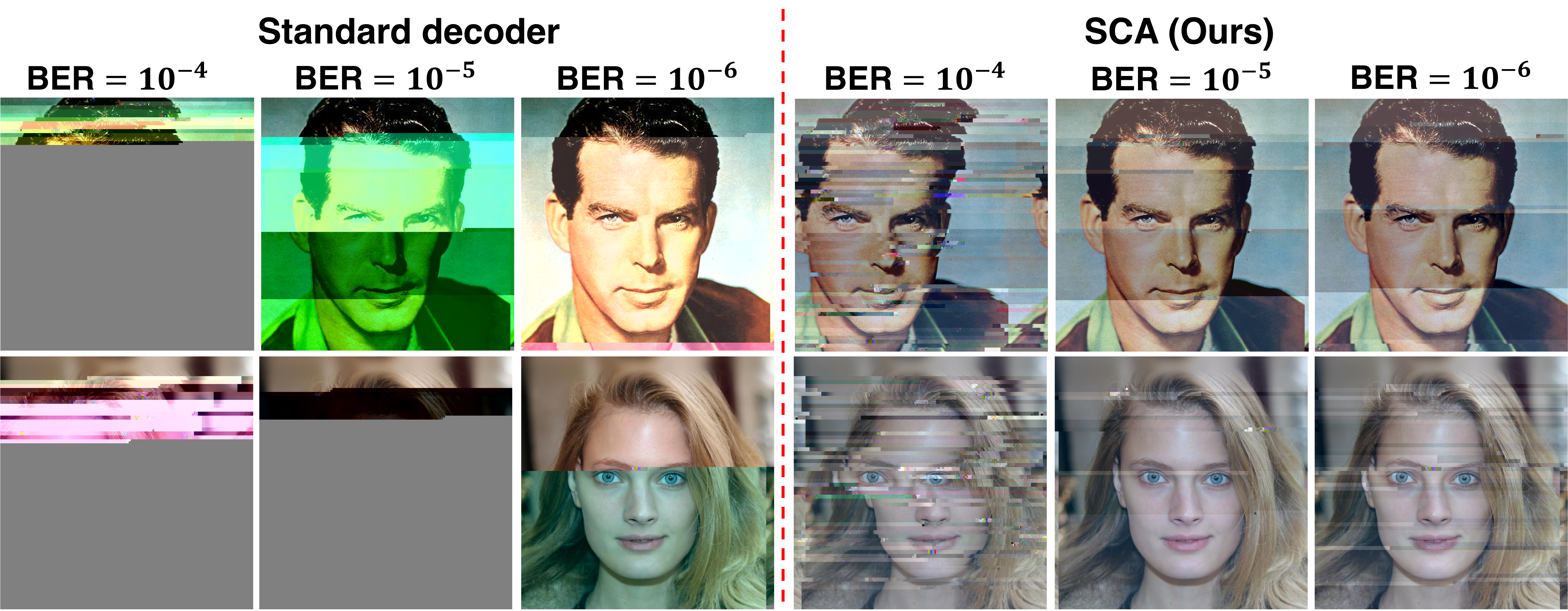}  
\captionsetup{font=footnotesize}
\vspace{-0.05in}
\caption{Visual comparison of the standard decoder's results (Left) with our SCA's results (Right) on various BERs.} 
\label{f:BER_b2}
\end{figure}

\textbf{Varying BERs of degraded images.} We also evaluate our method in more complex datasets, i.e., AFHQ~\cite{choi2020stargan} and CelebA-HQ~\cite{liu2015faceattributes}, and test the generalization of our method in handling varying degrees of BERs in the JPEG file without retraining, including BERs on the encrypted bitstream of $10^{-6}$, $10^{-5}$, and $10^{-4}$. The training is under the BER$=10^{-5}$, and the testing is under various BERs. Experimental results are shown in Tab.~\ref{t:BER}. We can observe that our two-stage method can still achieve outstanding results ($>$32dB) in large BER of $10^{-4}$.



Moreover, qualitative comparisons on CelebA-HQ with different BERs are shown in Fig.~\ref{f:BER_detail} and Fig.~\ref{f:BER_b2}. The left part of Fig.~\ref{f:BER_b2} shows the bitstream-corrupted JPEG images with varying degrees of BERs decoded by the standard decoder. 
Compared to the standard decoder, the proposed SCA shows great ability in resolving this problem, which can deliver much better visual results as BER decreases. In Fig.~\ref{f:BER_detail}, our method achieves the best results with more textures at the BER of $10^{-6}$, and even if the BER is up to $10^{-4}$, our method still can guarantee a good visual result compared to the bicubic upsampling method. These Results demonstrate the generalization ability of our method. We provide more visual results in the Supplementary Materials. 

%

\begin{table}[t]
\centering
\vspace{0.05in}
\captionsetup{font=footnotesize}
\caption{Quantitative comparison on varying BERs in terms of PSNR/SSIM with the training under BER$=10^{-5}$} 
\label{t:BER}
\vspace{-0.1in}
\footnotesize
\begin{tabular}{cccc} \hline 
Method  & BER  &  AFHQ~\cite{choi2020stargan} & CelebA-HQ~\cite{liu2015faceattributes} \\  
\hline 
\multirow{3}{*}{\shortstack{SCA \\ (Ours)}} & $10^{-4}$ &  13.44/0.45 & 12.29/0.57  \\  
 & $10^{-5}$  & 16.77/0.70 & \textbf{16.06/0.77} \\
 & $10^{-6}$  & \textbf{17.46/0.73} & 15.82/0.74 \\ \hdashline
\multirow{3}{*}{\shortstack{SCA+GCA \\ (Ours)}}   & $10^{-4}$  & 32.41/0.86 & 35.45/0.92 \\
 & $10^{-5}$ & 39.00/0.94  & 41.76/0.97 \\
 & $10^{-6}$ & \textbf{41.39/0.95}  & \textbf{41.85/0.97} \\
\hline          
\end{tabular}

\end{table}

\vspace{-0.1in}
\section{Conclusion}
\vspace{-0.05in}
This paper introduced a real-world JPEG image restoration problem with bit errors on the encrypted bitstream. We proposed a robust JPEG decoder, followed by a two-stage compensation and alignment work to restore bitstream-corrupted JPEG images. The robust JPEG decoder adopts an error-resilient mechanism to decode the corrupted JPEG bitstream. The two-stage framework comprises the self-compensation and alignment (SCA) stage and the guided compensation and alignment (GCA) stage, which aims to resolve the decoded images' color cast and block shift problem. The SCA is based on image content similarity free from training data, and the GCA employs coarse and refine-guided networks to restore full-resolution images gradually. Extensive experimental results and ablation studies show the effectiveness of our proposed method. We believe that this problem and our solution have the potential to be further explored in future studies.


\vspace{-0.1in}
\section*{Acknowledgement}
\vspace{-0.05in}
This research/project is supported by the National Research Foundation, Singapore, and Cyber Security Agency of Singapore under its National Cybersecurity R\&D Programme (NRF2018NCR-NCR009-0001). Any opinions, findings and conclusions or recommendations expressed in this material are those of the author(s) and do not reflect the views of National Research Foundation, Singapore and Cyber Security Agency of Singapore.



{\small
\bibliographystyle{ieee_fullname}
\bibliography{egbib}
}

\end{document}


\title{Supplementary Material---Bitstream-Corrupted JPEG Images are Restorable: Two-stage Compensation and Alignment Framework for Image Restoration\vspace{-1.5em}}


\author[1]{Wenyang Liu\thanks{$^\ast$Corresponding authors}}
\author[1$\ast$]{Yi Wang}
\author[1$\ast$]{Kim-Hui Yap}
\author[2]{Lap-Pui Chau\vspace{-0.8em}} 
\affil[1]{\textit {School of Electrical and Electronics Engineering, Nanyang Technological University, Singapore}}
\affil[2]{\textit {Dept. of Electronic and Information Engineering, The Hong Kong Polytechnic University, Hong Kong}}
\affil[ ]{\tt \small  {\{wenyang001, wang1241\}@e.ntu.edu.sg, ekhyap@ntu.edu.sg, lap-pui.chau@polyu.edu.hk}\vspace{-1.5em}}

\renewcommand\Authands{ and }

\maketitle

This document supplies more detailed information and visual comparisons of our method. We first introduce the employed pix2pix network and loss function in Section 3.3 of the manuscript. Then, we show more visual results in different experimental conditions.

\begin{figure*}[t]
\centering
\vspace{-0.2in}
\includegraphics[width=6.8in]{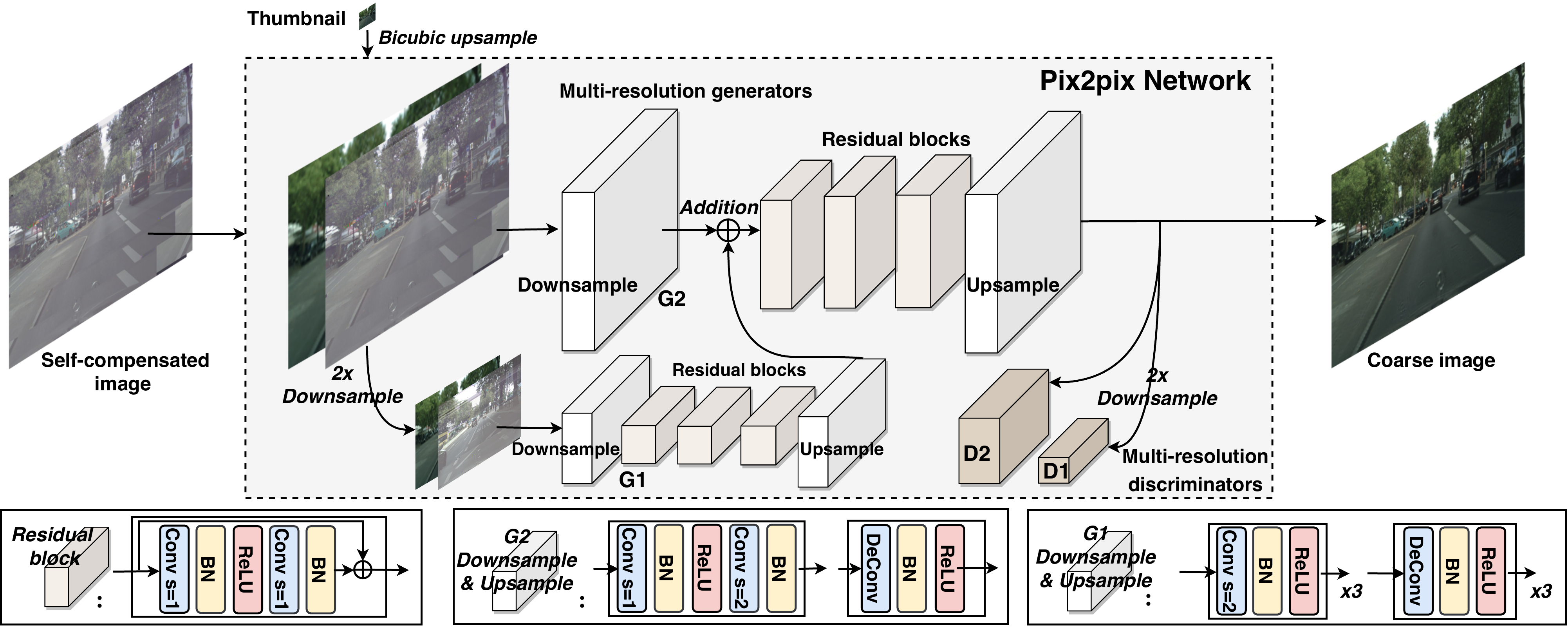}
\captionsetup{font=footnotesize}
\caption{Architecture of the pix2pix network. The input consists of two images: the self-compensated image (from the self-compensation and alignment (SCA) stage) and the extracted thumbnail (from the JPEG file's header). The output is the coarse image, which is guided by the thumbnail. The coarse image is then sent to a refine-guided Laplacian pyramid fusion network to refine details (see Figure 2 of the manuscript). The details of each component are shown at the bottom of the figure. $s$ means the stride of the convolution.} 
\label{f:network}
\vspace{-0.2in}
\end{figure*}

\section{Pix2pix Network}
Pix2pix~\cite{isola2017image, wang2018high, qu2019enhanced} networks were widely used to achieve an image-to-image translation. In this paper, the guided compensation and alignment (GCA) stage casts this image restoration problem as a task of image-to-image translation under the guidance of the extracted thumbnail. The employed pix2pix network can be regarded as a coarse-guided restoration network in the GCA. The architecture of the pix2pix network is shown in Fig.~\ref{f:network} which consists of multi-resolution generators (i.e., $G1$ and $G2$) and multi-resolution discriminators (i.e., $D1$ and $D2$). The aim is to generate a coarsely color-compensated and aligned image by fusing the self-compensated image (from the self-compensation and alignment (SCA) stage) and the bicubic upsampled low-resolution thumbnail (from the JPEG file's header).

\textbf{Multi-resolution generators.} The multi-resolution generators consist of two generators: a global generator $G1$ and a local generator $G2$ as shown in Fig.~\ref{f:network}. Both generators consist of a convolutional downsampling front-end, three residual blocks, and a transposed convolutional upsampling back-end, where the details of each component are shown at the bottom of Fig.~\ref{f:network}. The input of $G2$ is the concatenated images of the self-compensated image and the upsampled thumbnail, and the input of $G1$ is 2x downsampled images from the input of $G2$. The corresponding output of $G1$ is element-wise added with the feature maps from the downsampling front-end of $G2$. Ref.~\cite{wang2018high} proved that the multi-resolution generator structure is able to effectively integrate the learned global and local information from the image inputs to generate high-resolution image synthesis. In our paper, the global information of the upsampled thumbnail, i.e., color and structure information, can coarsely and implicitly guide the compensation and alignment of the self-compensated image that suffers from color cast and block shifts. The final high-resolution image is restored in structure and color except for realistic details, which will be sent to a refine-guided Laplacian pyramid fusion network to refine details (see Figure 2 of the manuscript).





\textbf{Multi-resolution discriminators.} The multi-resolution discriminators contains two discriminators $D1$ and $D2$, which have an identical architecture. The real and synthesized high-resolution images are downsampled by a factor of 2, such that the two-scale real and synthesized images are employed to train $D1$ (with low scale) and $D2$ (with high scale), respectively. The multi-resolution discriminators encourage the generators to produce both globally and locally consistent images with different scales of the receptive field.

\section{Loss Function}
Here we introduce the adversarial loss $L_{A}$, the feature matching loss $L_{FM}$, and the perceptual loss $L_{VGG}$ in Eq. (6) of the manuscript.

\textbf{Adversarial loss.} The adversarial loss is defined as a multi-task learning loss:

\begin{equation}\label{e:LA}
L_A = \min _G \max _{D_1, D_2} \sum_{k=1,2} \mathcal{L}_{\mathrm{GAN}}\left(G, D_k\right)
\end{equation}

\noindent where $\mathcal{L}_{\mathrm{GAN}}\left(G, D_k\right)$ is the adversarial loss of the k-the discriminator of $D_k$, expressed as:

\begin{equation}\label{e:LA1}
\mathcal{L}_{\mathrm{GAN}}\left(G, D_k\right) = E_{(X)}\text{log}D_k(X) + E_{(X)}\text{log}D_k(G(X_s, T)) 
\end{equation}

\noindent where $X$, $X_s$, and $T$ denote error-free real images, self-compensated images, and the extracted thumbnail. $G(X_s, T)$ represents the output by the pix2pix's generator.

\textbf{Feature matching loss.} Feature matching loss is defined as the matching similarity of features in multiple layers of the discriminator between the error-free real images and the generated images, expressed as:

\begin{equation}\label{e:FM}
L_{FM} = \min _G \sum_{k=1,2} \mathcal{L}_{FM}\left(G, D_k\right)
\end{equation}

\noindent where $\mathcal{L}_{FM}\left(G, D_k\right)$ is the feature matching loss with the k-th discriminator $D_k$, expressed as:

\begin{equation}\label{e:FM2}
L_{FM} = E_{(X)} \sum_{i=1}^{L} \frac{1}{N_i}\left[ \|D_k^{(i)}(X)-D_k^{(i)}(G(X_s, T))\|_1 \right]
\end{equation}

\noindent where $L$ is the total number of layers used for feature extraction, $N_i$ denotes the
number of elements in the $i$-th layer, $D_k^{(i)}$ is the extracted feature maps of the i-th layer in $D_k$.

\textbf{Perceptual loss.} Perceptual loss is used to measure the high-level differences, e.g., content and style discrepancies, between images. It is defined by the differences between pre-trained VGGNet extracted feature maps, expressed as:

\begin{equation}\label{e:PL}
L_{VGG}^{\phi, i}(\hat{X}, X)=\frac{1}{C_iH_iW_i}\|\phi_i(\hat{X})-\phi_i(X)\|_1
\end{equation}

\noindent where $\hat{X}$ is the final restored image of the network, $H_i$, $W_i$, and $C_i$ are the height, width, and channel of the $i$-th layer in VGGNet. $\phi_i()$ denotes the output feature map of the $i$-th layer.

\section{More Visual Results}
\textbf{Comparison of SCA with/without alignment.} Fig.~\ref{f:sp_a} shows a visual comparison of SCA with/ without block alignment processing. As we can see from the figure, although the proposed block alignment processing does not make the processed image fully aligned, this processing delivers better-aligned results than the SCA without the alignment, which proves its effectiveness.

\textbf{Comparison of coarse and refined images.} Fig.~\ref{f:sp6} shows a visual comparison of the coarse images by the pix2pix network and the refined images by the Laplacian fusion network. We can observe that the coarse images are not fully aligned and have some artifacts. After the refine-guided Laplacian fusion network processing, these artifacts are removed, and more texture details are generated, which proves the effectiveness of the proposed Laplacian fusion network.

\textbf{Comparison of other methods.} Fig.~\ref{f:sp1} shows a 1k-resolution visual comparison of our robust decoder, SCA, and GCA methods with standard decoder and the EPDN~\cite{qu2019enhanced} method. Figs.~\ref{f:sp2} and \ref{f:sp5} show a 2k-resolution visual comparison. We can see that our method consistently  has superior results over other methods in different-resolution image restoration.

\textbf{Generalization of varying BERs of images.} Fig.~\ref{f:sp4} shows more visual comparisons of the standard decoder, our SCA, and our SCA+GCA (two-stage model) on the AFHQ~\cite{choi2020stargan} dataset with different bit error rates (BERs). The training is under the BER=$10^{-5}$, and the testing is under various BERs. These results demonstrate the superior generalization ability of our two-stage method in handling varying degrees of BERs of the JPEG file without retraining.

{\small
\bibliographystyle{ieee_fullname}
\bibliography{egbib}
}

\begin{figure*}[t]
\centering
\includegraphics[width=5.7in]{supp/Sp_align_compressed.pdf}  
\captionsetup{font=footnotesize}
\caption{Visual comparison of the SCA with/without the block alignment processing on the 2k-resolution Cityscape~\cite{cordts2016cityscapes} dataset.}
\label{f:sp_a}
\end{figure*}

\begin{figure*}[b]
\centering
\includegraphics[width=5.0in]{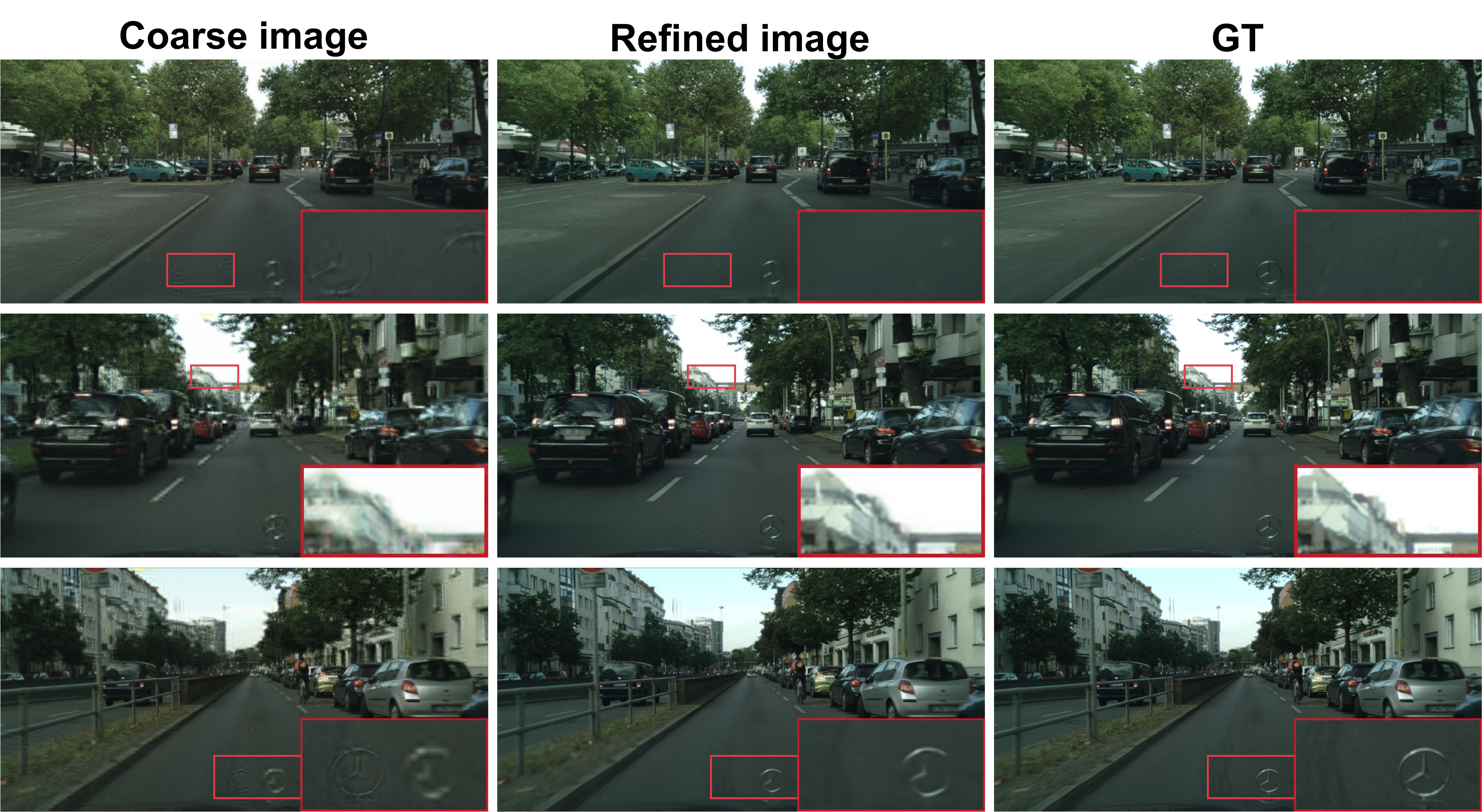}  
\captionsetup{font=footnotesize}
\vspace{-0.1in}
\caption{Visual comparison between coarse images obtained by the pix2pix network and refined images obtained by the Laplacian fusion network on 1k-resolution Cityscape~\cite{cordts2016cityscapes} dataset.}
\vspace{-0.1in}
\label{f:sp6}
\end{figure*}

\begin{figure*}[t]
\centering
\includegraphics[width=5.0in]{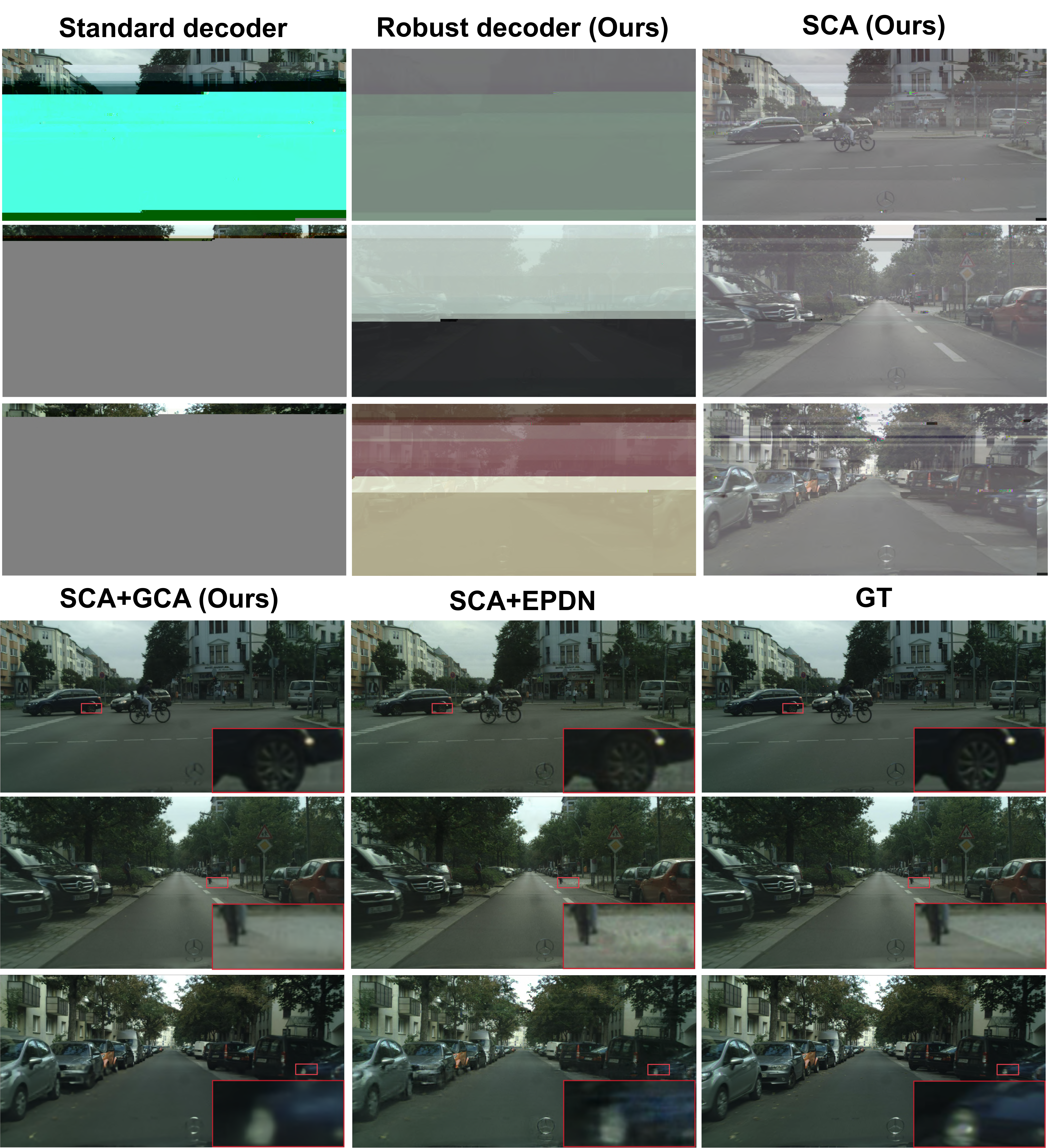}  
\captionsetup{font=footnotesize}
\vspace{-0.1in}
\caption{Visual comparison of the proposed robust decoder, SCA, and GCA methods with the standard decoder and
the EPDN~\cite{qu2019enhanced} method on 1k-resolution Cityscape~\cite{cordts2016cityscapes} dataset.} 
\vspace{-0.1in}
\label{f:sp1}
\end{figure*}

\begin{figure*}[htbp!]
\centering
\includegraphics[width=6.7in]{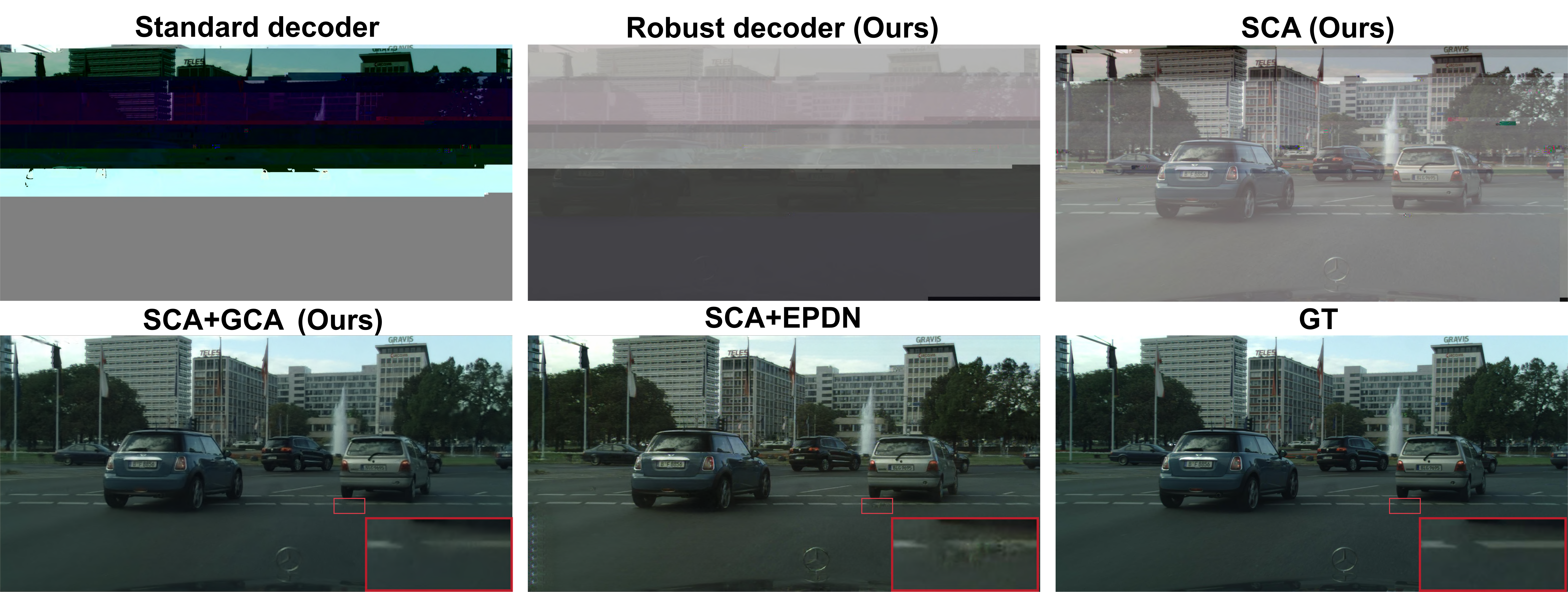}  
\captionsetup{font=footnotesize}
\vspace{-0.1in}
\caption{Visual comparison of the proposed robust decoder, SCA, and GCA methods with the standard decoder and
the EPDN~\cite{qu2019enhanced} method on 2k-resolution Cityscape~\cite{cordts2016cityscapes} dataset.} 
\vspace{-0.1in}
\label{f:sp2}
\end{figure*}

\begin{figure*}[htbp!]
\centering
\includegraphics[width=6.7in]{supp/Sp_5_compressed.pdf}  
\captionsetup{font=footnotesize}
\vspace{-0.1in}
\caption{Visual comparison of the proposed robust decoder, SCA, and GCA methods with the standard decoder and
the EPDN~\cite{qu2019enhanced} method on 2k-resolution Cityscape~\cite{cordts2016cityscapes} dataset.} 
\vspace{-0.1in}
\label{f:sp5}
\end{figure*}

\begin{figure*}[htbp!]
\centering
\includegraphics[width=6.7in]{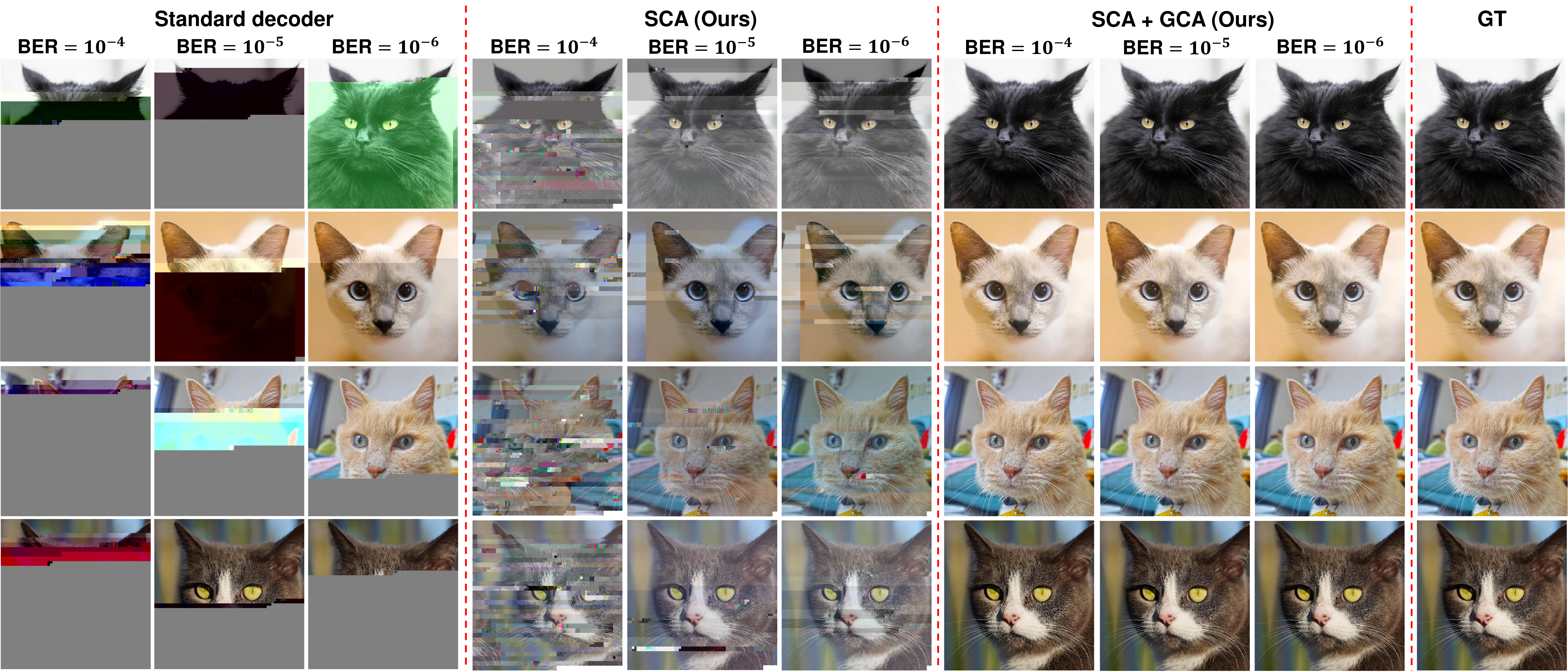}  
\captionsetup{font=footnotesize}
\vspace{-0.1in}
\caption{Visual comparison of the standard decoder’s results (Left) with
our SCA’s (middle) and SCA+GCA's results (Right) on various BERs in the AFHQ~\cite{choi2020stargan} dataset.}
\vspace{-0.1in}
\label{f:sp4}
\end{figure*}